\documentclass[reqno,12pt,a4paper]{amsart}

\usepackage{mathrsfs}
\usepackage{amssymb}
\usepackage{amsfonts}
\usepackage{latexsym}
\usepackage{amsthm}

\usepackage{graphicx}
\def\lb{\label}

\newcommand{\er}[1]{\textrm{(\ref{#1})}}

\begin{document}


\renewcommand{\theequation}{\arabic{section}.\arabic{equation}}
\theoremstyle{plain}
\newtheorem{theorem}{\bf Theorem}[section]
\newtheorem{lemma}[theorem]{\bf Lemma}
\newtheorem{corollary}[theorem]{\bf Corollary}
\newtheorem{proposition}[theorem]{\bf Proposition}
\newtheorem{definition}[theorem]{\bf Definition}
\newtheorem{remark}[theorem]{\it Remark}

\def\a{\alpha}  \def\cA{{\mathcal A}}     \def\bA{{\bf A}}  \def\mA{{\mathscr A}}
\def\b{\beta}   \def\cB{{\mathcal B}}     \def\bB{{\bf B}}  \def\mB{{\mathscr B}}
\def\g{\gamma}  \def\cC{{\mathcal C}}     \def\bC{{\bf C}}  \def\mC{{\mathscr C}}
\def\G{\Gamma}  \def\cD{{\mathcal D}}     \def\bD{{\bf D}}  \def\mD{{\mathscr D}}
\def\d{\delta}  \def\cE{{\mathcal E}}     \def\bE{{\bf E}}  \def\mE{{\mathscr E}}
\def\D{\Delta}  \def\cF{{\mathcal F}}     \def\bF{{\bf F}}  \def\mF{{\mathscr F}}
\def\c{\chi}    \def\cG{{\mathcal G}}     \def\bG{{\bf G}}  \def\mG{{\mathscr G}}
\def\z{\zeta}   \def\cH{{\mathcal H}}     \def\bH{{\bf H}}  \def\mH{{\mathscr H}}
\def\e{\eta}    \def\cI{{\mathcal I}}     \def\bI{{\bf I}}  \def\mI{{\mathscr I}}
\def\p{\psi}    \def\cJ{{\mathcal J}}     \def\bJ{{\bf J}}  \def\mJ{{\mathscr J}}
\def\vT{\Theta} \def\cK{{\mathcal K}}     \def\bK{{\bf K}}  \def\mK{{\mathscr K}}
\def\k{\kappa}  \def\cL{{\mathcal L}}     \def\bL{{\bf L}}  \def\mL{{\mathscr L}}
\def\l{\lambda} \def\cM{{\mathcal M}}     \def\bM{{\bf M}}  \def\mM{{\mathscr M}}
\def\L{\Lambda} \def\cN{{\mathcal N}}     \def\bN{{\bf N}}  \def\mN{{\mathscr N}}
\def\m{\mu}     \def\cO{{\mathcal O}}     \def\bO{{\bf O}}  \def\mO{{\mathscr O}}
\def\n{\nu}     \def\cP{{\mathcal P}}     \def\bP{{\bf P}}  \def\mP{{\mathscr P}}
\def\r{\rho}    \def\cQ{{\mathcal Q}}     \def\bQ{{\bf Q}}  \def\mQ{{\mathscr Q}}
\def\s{\sigma}  \def\cR{{\mathcal R}}     \def\bR{{\bf R}}  \def\mR{{\mathscr R}}
\def\S{\Sigma}  \def\cS{{\mathcal S}}     \def\bS{{\bf S}}  \def\mS{{\mathscr S}}
\def\t{\tau}    \def\cT{{\mathcal T}}     \def\bT{{\bf T}}  \def\mT{{\mathscr T}}
\def\f{\phi}    \def\cU{{\mathcal U}}     \def\bU{{\bf U}}  \def\mU{{\mathscr U}}
\def\F{\Phi}    \def\cV{{\mathcal V}}     \def\bV{{\bf V}}  \def\mV{{\mathscr V}}
\def\P{\Psi}    \def\cW{{\mathcal W}}     \def\bW{{\bf W}}  \def\mW{{\mathscr W}}
\def\o{\omega}  \def\cX{{\mathcal X}}     \def\bX{{\bf X}}  \def\mX{{\mathscr X}}
\def\x{\xi}     \def\cY{{\mathcal Y}}     \def\bY{{\bf Y}}  \def\mY{{\mathscr Y}}
\def\X{\Xi}     \def\cZ{{\mathcal Z}}     \def\bZ{{\bf Z}}  \def\mZ{{\mathscr Z}}
\def\O{\Omega}
\def\th{\theta}

\newcommand{\gA}{\mathfrak{A}}
\newcommand{\gB}{\mathfrak{B}}
\newcommand{\gC}{\mathfrak{C}}
\newcommand{\gD}{\mathfrak{D}}
\newcommand{\gE}{\mathfrak{E}}
\newcommand{\gF}{\mathfrak{F}}
\newcommand{\gG}{\mathfrak{G}}
\newcommand{\gH}{\mathfrak{H}}
\newcommand{\gI}{\mathfrak{I}}
\newcommand{\gJ}{\mathfrak{J}}
\newcommand{\gK}{\mathfrak{K}}
\newcommand{\gL}{\mathfrak{L}}
\newcommand{\gM}{\mathfrak{M}}
\newcommand{\gN}{\mathfrak{N}}
\newcommand{\gO}{\mathfrak{O}}
\newcommand{\gP}{\mathfrak{P}}
\newcommand{\gQ}{\mathfrak{Q}}
\newcommand{\gR}{\mathfrak{R}}
\newcommand{\gS}{\mathfrak{S}}
\newcommand{\gT}{\mathfrak{T}}
\newcommand{\gU}{\mathfrak{U}}
\newcommand{\gV}{\mathfrak{V}}
\newcommand{\gW}{\mathfrak{W}}
\newcommand{\gX}{\mathfrak{X}}
\newcommand{\gY}{\mathfrak{Y}}
\newcommand{\gZ}{\mathfrak{Z}}

\def\ve{\varepsilon}   \def\vt{\vartheta}    \def\vp{\varphi}    \def\vk{\varkappa}

\def\Z{{\mathbb Z}}    \def\R{{\mathbb R}}   \def\C{{\mathbb C}}    \def\K{{\mathbb K}}
\def\T{{\mathbb T}}    \def\N{{\mathbb N}}   \def\dD{{\mathbb D}}


\def\la{\leftarrow}              \def\ra{\rightarrow}            \def\Ra{\Rightarrow}
\def\ua{\uparrow}                \def\da{\downarrow}
\def\lra{\leftrightarrow}        \def\Lra{\Leftrightarrow}


\def\lt{\biggl}                  \def\rt{\biggr}
\def\ol{\overline}               \def\wt{\widetilde}
\def\no{\noindent}


\let\ge\geqslant                 \let\le\leqslant
\def\lan{\langle}                \def\ran{\rangle}
\def\/{\over}                    \def\iy{\infty}
\def\sm{\setminus}               \def\es{\emptyset}
\def\ss{\subset}                 \def\ts{\times}
\def\pa{\partial}                \def\os{\oplus}
\def\om{\ominus}                 \def\ev{\equiv}
\def\iint{\int\!\!\!\int}        \def\iintt{\mathop{\int\!\!\int\!\!\dots\!\!\int}\limits}
\def\el2{\ell^{\,2}}             \def\1{1\!\!1}
\def\sh{\sharp}
\def\wh{\widehat}
\def\bs{\backslash}

\def\sh{\mathop{\mathrm{sh}}\nolimits}
\def\Area{\mathop{\mathrm{Area}}\nolimits}
\def\arg{\mathop{\mathrm{arg}}\nolimits}
\def\const{\mathop{\mathrm{const}}\nolimits}
\def\det{\mathop{\mathrm{det}}\nolimits}
\def\diag{\mathop{\mathrm{diag}}\nolimits}
\def\diam{\mathop{\mathrm{diam}}\nolimits}
\def\dim{\mathop{\mathrm{dim}}\nolimits}
\def\dist{\mathop{\mathrm{dist}}\nolimits}
\def\Im{\mathop{\mathrm{Im}}\nolimits}
\def\Iso{\mathop{\mathrm{Iso}}\nolimits}
\def\Ker{\mathop{\mathrm{Ker}}\nolimits}
\def\Lip{\mathop{\mathrm{Lip}}\nolimits}
\def\rank{\mathop{\mathrm{rank}}\limits}
\def\Ran{\mathop{\mathrm{Ran}}\nolimits}
\def\Re{\mathop{\mathrm{Re}}\nolimits}
\def\Res{\mathop{\mathrm{Res}}\nolimits}
\def\res{\mathop{\mathrm{res}}\limits}
\def\sign{\mathop{\mathrm{sign}}\nolimits}
\def\span{\mathop{\mathrm{span}}\nolimits}
\def\supp{\mathop{\mathrm{supp}}\nolimits}
\def\Tr{\mathop{\mathrm{Tr}}\nolimits}
\def\BBox{\hspace{1mm}\vrule height6pt width5.5pt depth0pt \hspace{6pt}}
\def\as{\text{as}}
\def\all{\text{all}}
\def\where{\text{where}}
\def\Dom{\mathop{\mathrm{Dom}}\nolimits}
\def\ch{\mathop{\mathrm{ch}}\nolimits}
\def\sh{\mathop{\mathrm{sh}}\nolimits}


\newcommand\nh[2]{\widehat{#1}\vphantom{#1}^{(#2)}}
\def\dia{\diamond}

\def\Oplus{\bigoplus\nolimits}



\def\qqq{\qquad}
\def\qq{\quad}
\let\ge\geqslant
\let\le\leqslant
\let\geq\geqslant
\let\leq\leqslant
\newcommand{\ca}{\begin{cases}}
\newcommand{\ac}{\end{cases}}
\newcommand{\ma}{\begin{pmatrix}}
\newcommand{\am}{\end{pmatrix}}
\renewcommand{\[}{\begin{equation}}
\renewcommand{\]}{\end{equation}}
\def\eq{\begin{equation}}
\def\qe{\end{equation}}
\def\[{\begin{equation}}
\def\bu{\bullet}
\newcommand{\fr}{\frac}
\newcommand{\tf}{\tfrac}

\title[{Resonances  of third order differential operators}]
{Resonances  of third order differential operators}

\date{\today}
\author[Evgeny Korotyaev]{Evgeny L. Korotyaev}
\address{Saint-Petersburg State University,
Universitetskaya nab. 7/9, St.
Petersburg, 199034, Russia, \ korotyaev@gmail.com, \
e.korotyaev@spbu.ru}


\begin{abstract}
\no We consider resonances for  third order ordinary differential operator
with compactly supported  coefficients  on the real line.
Resonance are defined as zeros of a Fredholm determinant on a non-physical
sheet of three sheeted Riemann surface. We determine upper bounds of the number
of resonances in complex discs at large radius.
We express the trace formula in terms of resonances only.

\end{abstract}

\maketitle


\section {Introduction and main results}
\setcounter{equation}{0}

\subsection{Introduction}
There are a lot of results about resonances for second order
operators.   Unfortunately, we do not know results about
resonances for high order operators. In this paper we begin to study
resonances for the case of third order ordinary differential
operators with compactly supported  coefficients  on the real line.

We consider the self-adjoint operator $H$ acting in $L^2(\R)$ and
given by
\[
\lb{Hpq} H=H_0+V, \qqq V=p\pa+\pa p+q,\qqq \pa=-i{d\/dx},
\]
where  the operator $H_0=\pa^3=i{d^3\/dx^3}$ is unperturbed.
We assume that the coefficients $(p,q)$ are compactly supported and  belong to the space $\mH=\mH_\g$ for some $\g>0$ defined by


\[
\mH=\big\{(p,q)\in L^2(\R)\os L^2(\R): p'\in L^2(\R),\qq
\supp (p,q)\in[0,\g]^2\big\}.
\]
Under the condition $(p,q)\in \mH$ the operator $V(H_0-i)^{-1}$ is compact, see Proposition  \ref{TGH}. Then the operators $H_0$ and $H$ are  self-adjoint on the same domain  and $C_0^\infty(\R)$ is a core for both $H_0$ and $H$.
 The spectrum of both $H_0$ and $H$ is purely absolutely continuous
 and covers the real line $\R$ (see \cite{C80, DTT82,B85, BDT88}).
In the present paper we consider the resonances of third order
differential operators with compactly  supported coefficients.
The resonances are defined as zeros of the Fredholm determinant $D$
on three sheeted Riemann surface of the spectral variable
$\l$. We obtain global estimates of the function
$D$ and determine asymptotics of the number of resonances
in the large disc.

Now we discuss the main problems when we study resonances for third order
differential operators. Recall that for second order operators the scattering amplitude
(or the Fredholm determinant, it is equivalent) is a basic function to
study resonances. Here the so-called Born term (the first term in the
scattering amplitude) is important and give a lot of information about
resonances, for any dimension.
For third  order operators the scattering amplitude
(or the Fredholm determinant, it is equivalent) is also the basic function to
study resonances. But here the Born term is a very simple function (see \er{A0}). Thus the Born term is not so useful and
 we have much more problems to study resonances.

Results about the operator $H$ (in general, here the coefficients
$p,q$ are not compactly supported) are used in the integration of the {\it bad Boussinesq equation} given by
\[
\lb{Be}
\ddot p={1\/3}\pa^2\Bigl(\pa^2 p+{4}p^2\Bigr),\qq\dot p=\pa q,
\]
here $ \dot u={\pa u\/\pa t}$ is the derivatives
with respect to the time  and $\pa u={\pa u\/\pa x}$ is the derivatives
with respect space variable (see \cite{BZ02}, \cite{DTT82}  and references therein).
It is equivalent to the Lax equation $\dot H=HK-KH$, where $K=-\pa^2+{4\/3}p$.

\subsection{Determinant}
In order to define the Fredholm determinant $D$ we rewrite $V$ in the form
\[
\lb{defVj}
\begin{aligned}
V=V_1V_2,\qq V_1=\c_{[0,\g]},\qqq
 V_2=  \big(2p\pa +(q-ip')\big),
\end{aligned}
\]
where $\c_A(x)=1$ on the set $A\ss \R$ and $\c_A(x)=0$ on the set
$\R\sm A$. Instead of spectral parameter $\l\in \C_\pm$ it is convenient to  introduce a new variable $k\in \K_\pm$ by $\l=k^3$. Here
 the corresponding domain $\K_\pm$ (the sector, see Fig \ref{fig1}) is defined by
$$
\textstyle \K_{\pm }= \rt\{k\in \C:\arg k\in
(0,\pm{\pi\/3})\rt\}\ss \C_\pm.
$$
The sector $\K_+$ of the variable $k$ corresponds to the upper
half-plane $\C_+$ of the spectral parameter $\l$. The sector $\K_-$
of the variable $k$ corresponds to the lower half-plane $\C_-$ of
the spectral parameter $\l$.
Define a free resolvent $R_0(k)$ and an operator $Y_\pm^0(k), k\in \K_\pm$ by
\[
\lb{Y0}
\begin{aligned}
 R_0(k)=(H_0-k^3)^{-1},\qq  Y_\pm^0(k)=V_2 R_0(k)V_1.
\end{aligned}
\]
We will show (see Lemma~\ref{TY}~i) that for each $(p,q)\in\mH$
the operator $Y_\pm^0(k),k\in \K_\pm,$ is trace class and analytic in $k$.
Thus we can define the Fredholm determinant $D_\pm$ by
\[
\label{a.2}
D_\pm(k)=\det (I +Y_\pm^0(k)),\qqq k\in \K_\pm,
\]
which is analytic in $\K_\pm$. Moreover, in Theorem~\ref{T1} we will
show that the function $D_\pm$ has an analytic extension from
$\K_\pm$ into the whole complex plane without zero. The determinant
$D_\pm$ has not zeros in $\ol\K_\pm$ , since the operator $H$ has
not eigenvalues. The zeros of the function $D_\pm$ in $\C$ are
called {\it resonances}. The basic properties
 of determinants (see below \er{2.2}, \er{B2}) give the identity
\[
\lb{DD} D_+(k)=\ol D_-(\ol k), \qqq \forall \ k\in \C\sm \{0\}.
\]
Due to this identity it is enough to consider $ D_+$ or $ D_-$.

\begin{theorem}
\lb{T1}
 Let $(p,q)\in\mH$. Then the determinant $D_\pm(k)$ is analytic in $\K_\pm$ and
has an analytic  extension from $\K_\pm$ into the whole complex
plane with a possible pole of order $m\le 3$ at zero. Furthermore, the
function $D_+(k)$ satisfies the following asymptotics and identity:
\[
\lb{asDK+} \log D_+(k)={2e_+ p_0\/3ik}+{e_-q_0\/3ik^2}+{O(1)\/k^3},
\]
as $|k| \to \infty$ uniformly in $\arg k\in[0,{\pi\/3}]$, where
$p_0=\int_\R p(x)dx$ and
\[
\lb{Tr1}   {1\/\pi}\int_\R \Im (e^{i{\pi\/3}}\log
D_+((\l+i0)^{1\/3})d\l={2\/3}p_0.
\]
\end{theorem}

\no {\bf Remark.}
 A proof of the asymptotics \er{asDK+} and the trace formula \er{Tr1} is standard,
 see e.g. \cite{Ko16}.

\medskip

\subsection{S-matrix}
We discuss S-matrix for the operators $H_0, H$.
It is well known that the wave operators $W_\pm=W_\pm(H,H_0)$ for
the pair $H_0, H$, given by
$$
W_\pm=s-\lim e^{itH}e^{-itH_0} \qqq \as \qqq t\to \pm\iy,
$$
exist and  are unitary (see \cite{DTT82,B85, BDT88}). Then the  scattering operator $\cS=W_+^*W_-$
is unitary. The operators $H_0$ and $\cS$ commute and thus are
simultaneously diagonalizable:
\[
\lb{DLH0}
L^2(\R)=\int_{\R}^\oplus \cH_\l d\l,\qqq
H_0=\int_{\R}^\oplus\l \1_\l d\l,\qqq \cS=\int_{\R}^\oplus
\cS(\l)d\l;
\]
here $\1_\l$ is the identity in the fiber space $\cH_\l=\C$ and
$\cS(\l)$
is the scattering matrix (a scalar function in our case) for the pair $H_0, H$.

\medskip

We describe an analytic extension of the S-matrix. We consider two
cases: firstly, an analytic extension of the S-matrix for positive energy $\cS(\l),
\l=k^3>0$; secondly an analytic extension of  the S-matrix for negative energy $\cS(\l),
\l=k^3<0$. We have the following results.

\begin{theorem}
\lb{TS} Let $(p,q)\in\mH$. Then

i) The S-matrix $S_+(k)=\cS(k^3), k>0$ has an analytic extension into the
sector $\K_+$ and a meromorphic extension into the whole complex
plane $\C$. The zeros of $S_+$ coincide with the zeros of $D_-$ and
the poles of $S_+$ are precisely the zeros of $D_+$ and $D_\pm$ satisfies
\[
\lb{DS1} D_-(k)=D_+(k)S_+(k) \qqq \forall \  k\in \C.
\]

ii) The S-matrix $S_-(k)=\cS(k^3), k\in e^{i{\pi\/3}}\R_+$ has an analytic
extension into the sector $\K_+$ and a meromorphic extension into
the whole complex plane $\C$. The zeros of $S_-$ coincide with the
zeros of $D_-(e^{-i2{\pi\/3}}k)$ and the poles of $S_-$ are
precisely the zeros of $D_+$ and $D_\pm$ satisfies
\[
\lb{DS2} D_-(e^{-i{2\pi\/3}}k)=D_+(k)S_-(k)\qqq \forall \  k\in \C.
\]
\end{theorem}

We remark that we define the resonances as zeros of the
Fredholm determinant $D_+$ of an analytic  continuation in the
Riemann surface, as usual. We can define the resonances as  the  poles of the
meromorphic continuation od the perturbed resolvent in the Riemann
surface. By Theorem \ref{TS}, the resonances equivalently are poles
of the scattering matrix or zeros of the determinant $D$.

\subsection{Estimates}

For $(p,q)\in \mH$ we introduce a radius $r_*\ge 1$ by
\[
\lb{Cr} \textstyle r_*=\max \{{4\/3}C_*,1\},\qqq  C_* =2\sqrt\g
\big(\|q-ip'\|_2+2\|p\|_2\big).
\]

We prove the main result of the present paper.

\begin{theorem}
\lb{T2}
 Let $(p,q)\in\mH$. Then the determinant $D_+(k)$  satisfies:
\[
\lb{esDK} |D_+(k)|\le 48e^{2\g |k|},\qqq\forall\qq |k|\ge r_*.
\]
\end{theorem}

\no {\bf Remark.} 1) The function $k^mD_+(k)$ is entire and \er{esN}
gives that this function is of exponential type, where $m$ is defined in Theorem \ref{T1}.
 We recall that an entire function $f(z)$ is
said to be of $exponential$ $ type$  if there is a constant $\a$
such that  $|f(z)|\leq $ const. $e^{\a |z|}$ everywhere (see \cite{Ko88}).

2) In the proof the following symmetry of the free resolvent is used. Let
$R_0^+(x-y,k), x,y\in \R, k\in \K_+$ be the kernel of the free
resolvent. This kernel is pure imaginary on the half-line
$e^{i{\pi\/6}}\R$. Thus it satisfies the basic identity
\[
\lb{sym2} R_0^+(e_ok,t)=-\ol R _0^ +(e_o \ol k,t)\qqq \forall \ k\ne
0,\qq e_o=e^{i{\pi\/6}},
\]
and the resolvent is symmetric with respect to the line $e_o\R$, see Fig.
\ref{fig1}.

3) We need also the well known result about the Hadamard factorization.
The function  $k^mD_+(k)$ is entire and denote by $(k_n)_{n=1}^{\iy} $ the sequence of its zeros $\neq 0$ (counted with multiplicity),
so arranged that $0<|k_1|\leq |k_2|\leq \dots$. Then due to \er{esDK}  we have
\[
\lb{HFa} D_+(k)={C\/k^m}e^{b k}\lim_{r\to \iy}\prod_{|k_n|\leq
r}\lt(1-{k\/k_n}\rt)e^{{k\/k_n}},\ \ \ \ C=\lim_{k\to 0}k^mD_+(k)\ne
0,
\]
for some $b\in \C$ and integer $m\in [0,3]$, where the product
converges uniformly in every bounded disc.

\medskip

Now we discuss estimates of the number of zeros of the function $D_+$
in the disc $|k|<r$ counted with multiplicity.


\begin{corollary}
\lb{TN}
 Let $(p,q)\in\mH$ and let $\cN(r)$
 be the number of zeros of the function $D_+$
in the disc $|k|<r$ counted with multiplicity  for $r\ge r_*$. Then
\[
\lb{esN}
\cN(r)\le (16+m\log 2)+{6\g \/\log 2}r,
\]
where  $m\le 3$ is the order of pole of $D_+$ at $k=0$.

\end{corollary}

\no {\bf Remark.} We have an upper estimate \er{esN}.  There is a problem to get a lower estimate.

\medskip

Our next corollary concerns the trace formula in terms of resonances.
Define the scattering phase $\f_{sc}$ by $S_+(k)=e^{-2i \f_{sc}(k)}, k>0$.
Since
$D_+(k)=1+o(1)$ as $|k|\to \iy, k\in \ol\K_+$ we define
\[
\lb{dfsc}
\log D_+(k)=\log |D_+(k)|+i\f_{sc}(k),\qqq  k\in
\ol\K_+
\]
by condition $\log D_+(k)=o(1)$ as $|k|\to \iy, k\in
\ol\K_+$.

\begin{corollary}
\lb{Ttr} Let  $(p,q)\in\mH$ and let $R(k)=(H-k^3)^{-1}, k\in \K_+$. Then  the function $\Tr\,
\big(R_0(k)-R(k)\big)$ is analytic in $\K_+$ and has  a meromorphic extension into the whole complex
plane. Moreover, the following identity (the trace formula) holds true:
\[
\lb{RR0}
3k^3 \Tr\,
\big(R_0(k)-R(k)\big)=-m+bk+k^2\sum_{n\ge 1}{1\/
k_n(k-k_n)},\qqq 
\]
 where  the series converges absolutely and uniformly on any
compact set of $\C\sm \{k_n, n\ge 1\}$,
and the scattering phase $\f_{sc}$ satisfies
\[
\lb{BW} \f_{sc}'(k)=\Im b+\sum_{n\ge
1}\rt({1\/|k_n-k|^2}-{1\/|k_n|^2}\rt)\Im k_n\qq \forall\ k>0,
\]
uniformly on any compact subset of ${\R_+}$.
\end{corollary}

{\bf Remark.}
 Note that the identity \er{BW}  is a Breit-Wigner type formula for resonances (see p. 53 of \cite{RS78}).

\subsection{Brief overview}
Concerning results on resonances, we recall that there are a lot of results about resonances   from a
physicists point of view, see \cite{LL65}, \cite{RS79}. Since then, properties of  resonances have been the
object of intense study and we refer to   \cite{SZ91}, \cite{Z99} for the
mathematical approach in the multi-dimensional case  and references
given there.

A lot of papers are devoted to resonances of the one-dimensional
Schr\"odinger operator, see Froese \cite{F97}, Hitrik \cite{H99}, Korotyaev \cite{K04},
Simon \cite{S00}, Zworski \cite{Z87} and references given there. We
recall that Zworski \cite{Z87} obtained the first results about the
asymptotic distribution of resonances for the Schr\"odinger operator
with compactly supported potentials on the real line (this result is
sharper than Theorem \ref{TN} in the present paper).
 Inverse problems (characterization, recovering,
uniqueness) in terms of resonances were solved by Korotyaev for a
Schr\"odinger operator with a compactly supported potential on the
real line \cite{K05} and the half-line \cite{K04}, see also Zworski
\cite{Z02}, Brown-Knowles-Weikard \cite{BKW03} concerning the
uniqueness.

The resonances for  one-dimensional operators
$-{d^2\/dx^2}+q_\pi+q$, where $q_\pi$ is periodic and $q$ is a
compactly supported potential were considered  by Firsova
\cite{F84}, Korotyaev \cite{K11}, Korotyaev-Schmidt \cite{KS12}.
Christiansen \cite{C06} considered resonances for steplike potentials.
 Lieb-Thirring inequality for the resonances was determined in
 \cite{K16}. The "local resonance" stability problems were considered in
\cite{Ko04}, \cite{MSW10}.

Inverse problem for n-th order differential operators on the unit interval
is discussed in  \cite{L66}, \cite{Y00}.
We mention that there are results about inverse problem for third order differential operators on the unit interval see \cite{A99}, \cite{A01}.
Spectral analysis and spectral asymptotics for the
third order operator with periodic coefficients are considered in  for \cite{BK14}, \cite{BK14}. Here the Riemann spectral surface has three sheets and the inverse
problem is still open.


\subsection{Plan of the paper}
In Section 2 we determine different properties of free resolvent
 and obtain basic estimates about $Y_\pm^0$.
Section 3 contains estimates and asymptotics of the Fredholm
determinant $D_\pm(k), k\in \K_\pm$ and the proof of Theorem \ref{T1}.
Section 4 establishes an analytic continuation of $D_\pm(k),k\in
\K_\pm$ and the meromorphic continuation of S-matrix into the complex plane.
In Section 5 we determine the global estimate of the Fredholm
determinant $D_+$ on the whole complex plane.
In Section 6 we prove the main theorems.
Appendix contains  estimates about the approximation of the scattering amplitude.

\section {Properties of the free resolvent}
\setcounter{equation}{0}

\subsection{The well-known facts}
By $\cB$ we denote the class of bounded operators.
Let $\cB_1$ and $\cB_2$ be the trace and
the Hilbert-Schmidt class equipped with the norm $\|\cdot \|_{\cB_1}$
and $ \|\cdot \|_{\cB_2}$ correspondingly.
We recall some well known facts. Let $A, B\in \cB$ and $AB, BA,X\in\cB_1$. Then
\[
\lb{2.1} \Tr AB=\Tr BA,
\]
\[
\lb{2.2} \det (I+ AB)=\det (I+BA),
\]
\[
\lb{B1} |\det (I+ X)|\le e^{\|X\|_{\cB_1}},
\]
\[
\lb{B2} \ol{\det (I+ X)}=\det (I+ X^*),
\]
see e.g., Sect.~3. in the book \cite{S05}. Let  the operator-valued
function $\O :\cD\to \cB_1$ be analytic for some domain $\cD\ss\C$
and $(I+\O(k))^{-1}\in \cB$ for any $k\in \cD$. Then for the
function $F(k)=\det (I+\O (k))$ we have
\[
\lb{2.3} F'(k)= F(k)\Tr (I+\O (k))^{-1}\O '(k),\qqq k\in \cD.
\]

\subsection{The free resolvent}
Define the Fourier transformation $\F: L^2(\R)\to L^2(\R)$ by
\[
\lb{Ft}
\begin{aligned}
\wh f(k)=(\F f)(k)={1\/\sqrt{2\pi}}\int_\R f(x)e^{-ik x}dx,\ \ \
k\in\R.
\end{aligned}
\]
\begin{figure}
\tiny
\unitlength 0.7mm 
\linethickness{0.4pt}
\ifx\plotpoint\undefined\newsavebox{\plotpoint}\fi 
\begin{picture}(139.75,114)(0,0)
\put(0,55.75){\line(1,0){139.75}}
\multiput(30.25,0)(.033726625111,.05053428317){2246}{\line(0,1){.05053428317}}
\multiput(32.5,114)(.033740039841,-.056274900398){2008}{\line(0,-1){.056274900398}}
\qbezier(73.5,64.75)(77.25,61.875)(77,55.5)
\qbezier(61.5,65.25)(67.125,69.25)(73.25,65.25)
\qbezier(60.75,66.5)(56.375,61.75)(56.5,56)
\put(81.25,60.){\makebox(0,0)[cc]{$\tf{\pi}{3}$}}
\put(68.25,71.75){\makebox(0,0)[cc]{$\tf{\pi}{3}$}}
\put(53.25,61){\makebox(0,0)[cc]{$\tf{\pi}{3}$}}
\multiput(4.18,19.93)(.0579521,.0325708){15}{\line(1,0){.0579521}}
\multiput(5.918,20.907)(.0579521,.0325708){15}{\line(1,0){.0579521}}
\multiput(7.657,21.884)(.0579521,.0325708){15}{\line(1,0){.0579521}}
\multiput(9.395,22.861)(.0579521,.0325708){15}{\line(1,0){.0579521}}
\multiput(11.134,23.838)(.0579521,.0325708){15}{\line(1,0){.0579521}}
\multiput(12.873,24.815)(.0579521,.0325708){15}{\line(1,0){.0579521}}
\multiput(14.611,25.792)(.0579521,.0325708){15}{\line(1,0){.0579521}}
\multiput(16.35,26.77)(.0579521,.0325708){15}{\line(1,0){.0579521}}
\multiput(18.088,27.747)(.0579521,.0325708){15}{\line(1,0){.0579521}}
\multiput(19.827,28.724)(.0579521,.0325708){15}{\line(1,0){.0579521}}
\multiput(21.565,29.701)(.0579521,.0325708){15}{\line(1,0){.0579521}}
\multiput(23.304,30.678)(.0579521,.0325708){15}{\line(1,0){.0579521}}
\multiput(25.042,31.655)(.0579521,.0325708){15}{\line(1,0){.0579521}}
\multiput(26.781,32.632)(.0579521,.0325708){15}{\line(1,0){.0579521}}
\multiput(28.52,33.609)(.0579521,.0325708){15}{\line(1,0){.0579521}}
\multiput(30.258,34.587)(.0579521,.0325708){15}{\line(1,0){.0579521}}
\multiput(31.997,35.564)(.0579521,.0325708){15}{\line(1,0){.0579521}}
\multiput(33.735,36.541)(.0579521,.0325708){15}{\line(1,0){.0579521}}
\multiput(35.474,37.518)(.0579521,.0325708){15}{\line(1,0){.0579521}}
\multiput(37.212,38.495)(.0579521,.0325708){15}{\line(1,0){.0579521}}
\multiput(38.951,39.472)(.0579521,.0325708){15}{\line(1,0){.0579521}}
\multiput(40.69,40.449)(.0579521,.0325708){15}{\line(1,0){.0579521}}
\multiput(42.428,41.426)(.0579521,.0325708){15}{\line(1,0){.0579521}}
\multiput(44.167,42.404)(.0579521,.0325708){15}{\line(1,0){.0579521}}
\multiput(45.905,43.381)(.0579521,.0325708){15}{\line(1,0){.0579521}}
\multiput(47.644,44.358)(.0579521,.0325708){15}{\line(1,0){.0579521}}
\multiput(49.382,45.335)(.0579521,.0325708){15}{\line(1,0){.0579521}}
\multiput(51.121,46.312)(.0579521,.0325708){15}{\line(1,0){.0579521}}
\multiput(52.859,47.289)(.0579521,.0325708){15}{\line(1,0){.0579521}}
\multiput(54.598,48.266)(.0579521,.0325708){15}{\line(1,0){.0579521}}
\multiput(56.337,49.243)(.0579521,.0325708){15}{\line(1,0){.0579521}}
\multiput(58.075,50.221)(.0579521,.0325708){15}{\line(1,0){.0579521}}
\multiput(59.814,51.198)(.0579521,.0325708){15}{\line(1,0){.0579521}}
\multiput(61.552,52.175)(.0579521,.0325708){15}{\line(1,0){.0579521}}
\multiput(63.291,53.152)(.0579521,.0325708){15}{\line(1,0){.0579521}}
\multiput(65.029,54.129)(.0579521,.0325708){15}{\line(1,0){.0579521}}
\multiput(66.768,55.106)(.0579521,.0325708){15}{\line(1,0){.0579521}}
\multiput(68.507,56.083)(.0579521,.0325708){15}{\line(1,0){.0579521}}
\multiput(70.245,57.06)(.0579521,.0325708){15}{\line(1,0){.0579521}}
\multiput(71.984,58.038)(.0579521,.0325708){15}{\line(1,0){.0579521}}
\multiput(73.722,59.015)(.0579521,.0325708){15}{\line(1,0){.0579521}}
\multiput(75.461,59.992)(.0579521,.0325708){15}{\line(1,0){.0579521}}
\multiput(77.199,60.969)(.0579521,.0325708){15}{\line(1,0){.0579521}}
\multiput(78.938,61.946)(.0579521,.0325708){15}{\line(1,0){.0579521}}
\multiput(80.676,62.923)(.0579521,.0325708){15}{\line(1,0){.0579521}}
\multiput(82.415,63.9)(.0579521,.0325708){15}{\line(1,0){.0579521}}
\multiput(84.154,64.877)(.0579521,.0325708){15}{\line(1,0){.0579521}}
\multiput(85.892,65.855)(.0579521,.0325708){15}{\line(1,0){.0579521}}
\multiput(87.631,66.832)(.0579521,.0325708){15}{\line(1,0){.0579521}}
\multiput(89.369,67.809)(.0579521,.0325708){15}{\line(1,0){.0579521}}
\multiput(91.108,68.786)(.0579521,.0325708){15}{\line(1,0){.0579521}}
\multiput(92.846,69.763)(.0579521,.0325708){15}{\line(1,0){.0579521}}
\multiput(94.585,70.74)(.0579521,.0325708){15}{\line(1,0){.0579521}}
\multiput(96.323,71.717)(.0579521,.0325708){15}{\line(1,0){.0579521}}
\multiput(98.062,72.694)(.0579521,.0325708){15}{\line(1,0){.0579521}}
\multiput(99.801,73.672)(.0579521,.0325708){15}{\line(1,0){.0579521}}
\multiput(101.539,74.649)(.0579521,.0325708){15}{\line(1,0){.0579521}}
\multiput(103.278,75.626)(.0579521,.0325708){15}{\line(1,0){.0579521}}
\multiput(105.016,76.603)(.0579521,.0325708){15}{\line(1,0){.0579521}}
\multiput(106.755,77.58)(.0579521,.0325708){15}{\line(1,0){.0579521}}
\multiput(108.493,78.557)(.0579521,.0325708){15}{\line(1,0){.0579521}}
\multiput(110.232,79.534)(.0579521,.0325708){15}{\line(1,0){.0579521}}
\multiput(111.971,80.511)(.0579521,.0325708){15}{\line(1,0){.0579521}}
\multiput(113.709,81.489)(.0579521,.0325708){15}{\line(1,0){.0579521}}
\multiput(115.448,82.466)(.0579521,.0325708){15}{\line(1,0){.0579521}}
\multiput(117.186,83.443)(.0579521,.0325708){15}{\line(1,0){.0579521}}
\multiput(118.925,84.42)(.0579521,.0325708){15}{\line(1,0){.0579521}}
\multiput(120.663,85.397)(.0579521,.0325708){15}{\line(1,0){.0579521}}
\multiput(122.402,86.374)(.0579521,.0325708){15}{\line(1,0){.0579521}}
\multiput(124.14,87.351)(.0579521,.0325708){15}{\line(1,0){.0579521}}
\multiput(125.879,88.328)(.0579521,.0325708){15}{\line(1,0){.0579521}}
\multiput(127.618,89.306)(.0579521,.0325708){15}{\line(1,0){.0579521}}
\multiput(129.356,90.283)(.0579521,.0325708){15}{\line(1,0){.0579521}}
\multiput(131.095,91.26)(.0579521,.0325708){15}{\line(1,0){.0579521}}
\multiput(132.833,92.237)(.0579521,.0325708){15}{\line(1,0){.0579521}}
\multiput(134.572,93.214)(.0579521,.0325708){15}{\line(1,0){.0579521}}
\multiput(136.31,94.191)(.0579521,.0325708){15}{\line(1,0){.0579521}}
\put(139.5,97.5){\makebox(0,0)[cc]{$e^{i{\pi\/6}}\R$}}
\put(125.75,81.75){\makebox(0,0)[cc]{$\K_+$}}
\put(66.25,108.25){\makebox(0,0)[cc]{$\K_+'$}}
\put(12.75,83.75){\makebox(0,0)[cc]{$\K_+''$}}
\put(11.75,18.25){\makebox(0,0)[cc]{$\K_-''$}}
\put(62.75,9.5){\makebox(0,0)[cc]{$\K_-'$}}
\put(119.25,21){\makebox(0,0)[cc]{$\K_-$}}
\end{picture}
\caption{\footnotesize The plane $\K$ is a union of sectors $\K_\pm, \K_\pm', \K_\pm''$ and $e^{i{\pi\/6}}\R$} is the symmetry line for $D_+$
\lb{fig1}
\end{figure}

We discuss the free  resolvent $R_0(k)=(H_0-k^3)^{-1}, k\in \K_{\pm }$.
The kernel $R_0(k,x-y), x,y\in \R$ of the free resolvent
$R_0(k), k\in \K_{\pm }$ is given by
\begin{equation}
\label{defG}
R_0(k,t)={1\/2\pi}\int_\R e^{i\x t}{d\x\/\x^3-k^3},\qq  t=x-y,\qq
\qq k\in \K_{\pm }.
\end{equation}
The equation $\x^3-k^3=0$ has three zeros:
$$
k_0=k,\qq k^\pm =ke_\pm,\qq  e_\pm=e^{\pm i{2\pi\/3}}.
$$
$\bu$ Consider the case $k\in \K_+$, i.e.,
when the spectral parameter $\l=k^3$ belongs to the upper
half-plane $\C_+$. We have for $t=x-y>0$:
$$
R_0^+(k,t)={1\/2\pi}\int_\R e^{i\x t}{d\x\/\x^3-k^3}= {2\pi
i\/2\pi}\rt({e^{itk}\/3k^2}+{e^{itk^+}\/3k^2e_+^2} \rt) =
{i\/3k^2}(e^{itk}+e_+e^{itk^+})
$$
and for $t=x-y<0$:
$$
R_0^+(k,t)={1\/2\pi}\int_\R e^{i\x t}{d\x\/\x^3-k^3}= -{2\pi
i\/2\pi}{1\/3k^2e_-^2}e^{itk^-} = -{ie_-\/3k^2}e^{itk^-}.
$$
We rewrite these identities in the form
\[
\begin{aligned}
\label{S3}
R_0^+(k,t)=
{i\/3k^2}\ca e^{itk}+e_+e^{itk^+}\qqq &  \as \qq t>0\\
-e_- e^{itk^-}\qqq  & \as \qq t<0 \ac,\qqq k\in \K_+.
\end{aligned}
\]
Recall that $\pa=-i{d\/dx}$.
Similar arguments give identities for the operator $\pa R_0(k)$:
\[
\begin{aligned}
\label{S4}
(\pa R_0^+)(k,t)=
{i\/3k}\ca (e^{itk}+e_- e^{itk^+})\qqq &  \as \qq t>0\\
-e_+ e^{itk^-}\qqq  & \as \qq t<0 \ac,\qqq k\in \K_+.
\end{aligned}
\]

$\bu$ Consider the case $k\in \K_{-}$, i.e.,
when the spectral parameter $\l=k^3$ belongs to the lower
half-plane $\C_-$. We have for $t=x-y<0$:
$$
R_0^-(k,t)={1\/2\pi}\int_\R e^{i\x t}{d\x\/\x^3-k^3}= -{2\pi
i\/2\pi}\rt({e^{itk}\/3k^2}+{e^{itk^-}\/3k^2e_-^2} \rt) =
-{i\/3k^2}(e^{itk}+e_-e^{itk^-})
$$
and for $t=x-y>0$:
$$
R_0^-(k,t)={1\/2\pi}\int_\R e^{i\x t}{d\x\/\x^3-k^3}= {2\pi
i\/2\pi}{1\/3k^2e_+^2}e^{itk^+} = {ie_+\/3k^2}e^{itk^+}.
$$
We rewrite these identities in the form
\[
\begin{aligned}
\label{S5}
R_0^-(k,t)=
{i\/3k^2}\ca e_+ e^{itk^+}\qqq  & \as \qq t>0 \\
(-e^{itk}-e_-e^{itk^-})\qqq &  \as \qq t<0 \ac,\qqq k\in \K_{-}.
\end{aligned}
\]
Similar arguments give following identities for the operator $\pa R_0(k)$:
\[
\begin{aligned}
\label{S4x}
(\pa R_0^-)(k,t)=
{i\/3k}\ca e_- e^{itk^+}\qqq &  \as \qq t>0\\
(-e^{itk}-e_+e^{itk^-}) \qqq  & \as \qq t<0 \ac,\qqq k\in \K_-.
\end{aligned}
\]

The definitions \er{defVj}, \er{Y0} imply
\[
\lb{exrepY0}
Y_\pm^0 =V_2 R_0(k)V_1 =\big(2p\pa +(q-ip')\big) R_0 V_1, \qqq k\in \K_\pm.
\]
Below we show that each operator $Y_\pm(k), k\in \K_\pm$ belongs to the
Hilbert Schmidt class and the mapping $Y_\pm :\K_\pm \to \cB_2$
is analytic. We have the following lemma.

\begin{proposition}
\label{TGH}
Let $(p,q)\in \mH$. Then

i) $Y_\pm^0(k)$ is  Hilbert Schmidt for all $k\in \C\sm \{0\}$ and analytic in $k$.

ii) the operator $VR_0(k)$ is  Hilbert Schmidt for all $k\in \K_\pm$.

iii) For each $t\in \R$  the kernel $R_0^+(k,t)$ is analytic in
$\C\sm \{0\}$ with pole at $k=0$ of order 2. Moreover, $i
R_0^+(k,t)$ is real on the line $e^{i{\pi\/6}}\R$ and

\end{proposition}

{\bf Proof.} i) Recall that  $Y_\pm^0(k)=V_2R_0(k)V_1, k\in \K_{\pm}$.
Let $Y_\pm^0(x,y,k)$ be its integral kernel. From explicit formulas
\er{S3} - \er{S4x} we deduce that   $Y_\pm^0(x,y,k)$ as a function of
$k$ has as analytic extension to the whole complex plane with pole
at $k=0$. From \er{S3} and \er{S4} we deduce that the operator
$Y_\pm^0(k), k\in \K_{\pm}$ has the kernel $Y_\pm^0(x,y,k)$  from
$L^2(\R^2)$  for all $k\in \C\sm \{0\}$, since the functions $p,q$
are compactly supported. The proof of ii) is similar.

iii)
For any
$k=re^{i{\pi\/6}}, r>0$ the identity \er{S3} gives
$$
\begin{aligned}
 iR_0^+(k,t)=
{1\/3r^2}\ca (\ol e_* e^{itk}+e_*e^{itk e_+})=2\Re (\ol e_* e^{itk}) \qqq &  \as \qq t>0\\
e^{itk e_-}=e^{tr}\qqq  & \as \qq t<0 \ac ,
\end{aligned}
$$
since $e_- =-e_*$ and $ke_+=-\ol k$ and $ike_-=r$. Thus
$iR_0^+(re^{i{\pi\/6}},t)$ is real for all $t$ and we have \er{sym2}.
 \BBox

In order to study resonances we  use  the symmetry \er{sym2}. Then it
is sufficient to study resonances only on the turned half plane $e^{i{\pi\/6}}\C_+$.

Using \er{S3} and \er{S5} we obtain the following formula
\[
R_0^+(k,t)-R_0^-(k,t)={i\/3k^2}e^{itk},\qqq k\ne 0, t=x-y\in \R.
\]
Thus we have the {\bf first} identity for $Y_\pm^0$:
\[
\lb{X} Y_+^0(k)-Y_-^0(k)=P(k),\qqq k\ne 0,
\]
where $P(k)$  is a rank one operator with a kernel $P(k,x,y)$ given
by
\[
\lb{Xxy} P(k,x,y)={i\/3k^2}V(x,k)
e^{i(x-y)k}\c_{[0,\g]}(y), \qq V(\cdot,k)=2kp+(q-ip').
\]
Moreover, we have
\[
\begin{aligned}
\label{Stx} R_0^-(k^-,t)=
{i\/3k^2e_-^2}\ca e_+ e^{itk}\\
-e^{itke_-}-e_-e^{itke_-^2}
\ac\!\!\!\!
=
{i\/3k^2}\ca  e^{itk}\   & \as \ \ t>0 \\
-e_-e^{itk^-}-e_+e^{itk^+}\  &  \as \ \ t<0 \ac.
\end{aligned}
\]
The relations \er{S3}, \er{Stx}  give
\[
\lb{E1} R_0^+(k,t)-R_0^-(k^-,t)={ie_+\/3k^2}e^{itke_+},\qqq k\ne 0,
t\in \R,
\]
which yields the {\bf second} identity for $Y_\pm^0$:
\[
\lb{E2} Y_+^0(k)-Y_-^0(k^-)=P(k^+),\qqq k\ne 0.
\]
 Thus from \er{X}-\er{E2} we obtain

\begin{proposition}\label{TYG}
If $k\ne 0$, then     $Y_+^0(k)-Y_-^0(k)$ and $Y_+^0(k)-Y_-^0(e_-k)$ are trace class and
satisfy
\[
\begin{aligned}
\label{S3x} \|Y_+^0(k)-Y_-^0(k)\|_{\cB_1}\le {\sqrt \g\/3|k|^2}
\|V(\cdot,k)\|_2 e^{\g |\Im k|},
\\
\|Y_+^0(k)-Y_-^0(k^-)\|_{\cB_1}\le {\sqrt \g\/3|k|^2} \|V(\cdot,k^-)\|_2
e^{\g |\Im k^-|}.
\end{aligned}
\]

\end{proposition}

\subsection{The operator valued function $Y_\pm^0$.}

In order to estimate $Y_\pm^0$ we need the following results.

\begin{lemma}
\label{TX} Let  $F_m(k)=a \pa^m R_0(k)b$ for $m=0,1$ and $k\in
\K_\pm$ for some $a,b\in L^2(\R)$.  Then

i) Each operator $F_0(k)\in \cB_j, j=1,2, k\in \K_\pm$; the
operator-valued function $X_0: \K_\pm\to \cB_j$ is analytic.
Moreover, $X_0(k)$ for all $k\in \K_\pm$  satisfies
\[
\lb{F2} \|F_0(k)\|_{\cB_2}\le {2\/3|k|^2}\|a\|_2 \|b\|_2,
\]
\[
\lb{F1}
\begin{aligned}
\|F_0(k)\|_{\cB_1}\le {\|a\|_2\|b\|_2\/\sqrt 3|k| (|\Im k| |\Im
k^+|)^{1\/2}}.
\end{aligned}
\]

ii) Each $F_1(k)\in\cB_1,k\in\K_\pm$ and  the operator-valued
function $F_1: \K_\pm\to \cB_1$ is analytic.
 Moreover, $X_1$ satisfies
\[
\lb{F12} \|F_1(k)\|_{\cB_2}\le {2\/3|k|}\|a\|_2 \|b\|_2,\qq
k\in\ol\K_\pm,
\]
\[
\lb{F11} \|F_1(k)\|_{\cB_1}\le {3\|a\|_2\|b\|_2\/2 (|\Im k| |\Im
k^+|)^{1\/2}}.
\]
\end{lemma}

{\bf Proof.} We consider $k\in \K_+$, the proof for $k\in \K_-$ is similar.
The estimates \er{S3} give
$$
\|F_0(k)\|_{\cB_2}^2\le {4\/9|k|^4} \int_\R dx \int_\R  |a(x)|^2
|b(y)|^2 dy= {4\/9|k|^4}\|a\|^2 \|b\|^2,
$$
which yields \er{F2}. Similar arguments and estimates \er{S3} imply
$$
\|F_1(k)\|_{\cB_2}^2\le {4\/9|k|^2} \int_\R dx \int_\R  |a(x)|^2
|b(y)|^2 dy= {4\/9|k|^2}\|a\|^2 \|b\|^2
$$
which yields \er{F12}. For $k\in \K_+$ we have
\[
\lb{2.4}
\begin{aligned}
{1\/\x^3-k^3}={1\/(\x-k)(\x-k^-)(\x-k^+)},\\
 k^\pm=e_\pm k,\qq
e_\pm=e^{\pm i{2\pi\/3}},\qqq k^+-k^-=ik{\sqrt 3},\\
-\Im k^-=\textstyle \sin ({\pi\/3}+\arg k)\in [1,{\sqrt 3\/2}].
\end{aligned}
\]
Using  the estimates \er{2.4}  and $|\Im k^-|\ge {\sqrt 3\/2}|k|$, we obtain
$$
\begin{aligned}
\|F_0(k)\|_{\cB_1}\le {1\/|\Im k^-|} \Big\|a
(\pa-k)^{-1}\Big\|_{\cB_2} \Big\|(\pa-k^+)^{-1}b \Big\|_{\cB_2}
\\
\le
{\|a\|_2\|b\|_2\/2\pi |\Im k^-|}\Big\|{1\/\x-k}\Big\|_2\Big\|{1\/\x+k^+}\Big\|_2=
{\|a\|_2\|b\|_2\/2|\Im k^-| (|\Im k| |\Im k^+|)^{1\/2}}\\
\le
{\|a\|_2\|b\|_2\/\sqrt 3|k| (|\Im k| |\Im k^+|)^{1\/2}},
\end{aligned}
$$
 since $\int_\R|s\pm k|^{-2}ds={\pi\/\Im k}$. Similar arguments imply
$$
\begin{aligned}
\|F_1(k)\|_{\cB_1}\le C_1 \Big\|a (\pa-k)^{-1}\Big\|_{\cB_2}
\Big\|(\pa-k^+)^{-1}b \Big\|_{\cB_2}\le {C_1\|a\|_2\|b\|_2\/2 (|\Im
k| |\Im k^+|)^{1\/2}},
\\
C_1=\sup_{\x\in \R} |g(\x)|,\qqq g(\x)= {|\x|\/|\x-k^-|} \le
1+{|k^-|\/|\x-k^-|}\le 1+{|k|\/|\Im k^-|}\le 1+{2\/\sqrt 3}\le 3
\end{aligned}
$$
which yields \er{F11} \BBox

Recall that for $(p,q)\in \mH$ we have introduced the constants
$$
 C_* =2\sqrt\g \big(\|q-ip'\|_2+2\|p\|_2\big),\qqq \textstyle
r_*=\max \{{4\/3}C_*,1\}, \qq V_0(k)=\int_\R V(x,k)dx.
$$

\begin{lemma}
\label{TY}
Let $(p,q)\in\mH$. Then
the operator $Y_\pm^0(k)\in \cB_j, j=1,2$ for each $k\in \K_\pm$,  the operator-valued
function $Y_\pm^0: \K_\pm\to \cB_j$ is analytic and has an analytic   extension into the whole complex plane without zero.
Moreover, $Y_\pm^0(k)$ satisfies for all $k\in\K_\pm$:
\[
\begin{aligned}
\lb{V1} |V_0(k)|\le \sqrt \g \|V(\cdot,k)\|_{2},\qq
\forall \ k,\\
\sqrt \g \|V(\cdot,k)\|_{2}\le |k|C_*, \qq \forall \ |k|\ge r_*,
\end{aligned}
\]
and
\[
\lb{Y2} \|Y_\pm^0 (k)\|_{\cB_2}\le {(1+|k|)\/3|k|^2}C_*,
\]
\[
\lb{Y1}
\begin{aligned}
\|Y_\pm^0(k)\|_{\cB_1}\le   {(1+|k|)\/|k| }{C_*\/ (|\Im k| |\Im
k^+|)^{1\/2}},
\end{aligned}
\]
and in particular,
\[
\lb{Y2c} \|Y_\pm^0 (k)\|_{\cB_2}\le {1\/2}, \qqq \ if \qqq |k|\ge r_*.
\]

\end{lemma}

{\bf Proof.} We have
$$
\begin{aligned}
|V_0|=|\int_\R(2p k+(q-ip'))dx|\le \sqrt \g \|V(\cdot,k)\|_{2},\qq
\forall \ k,
\\
{|V_0|\/|k|}\le {\sqrt \g\/|k|} \|V(\cdot,k)\|_{2}\le \sqrt \g
\|2p\|_{2}+\sqrt \g {\|q-ip'\|_2\/|k|}\le {C_*\/2}+{C_*\/2r_*}\le
C_* \qq \forall \ |k|\ge r_*.
\end{aligned}
$$

We rewrite $Y_\pm^0$ in the form
\[
\lb{YXT}
Y_\pm^0=X_{\pm}+T_\pm,\qq X_\pm=(q-ip')R_\pm^0(k)V_1,\qq T_\pm=2p\pa R_\pm^0(k)V_1.
\]
From Lemma \ref{TX} we deduce that
the mapping $Y_\pm^0:\K_\pm\to \cB_2$ is analytic and satisfies
$$
\|Y_\pm^0(k)\|_{\cB_2}\le\|X_\pm(k)\|_{\cB_2}+\|T_\pm(k)\|_{\cB_2}\le {2\/3|k|^2}\|q-ip'\|_2 \|V_1\|_2+{4\/3|k|}\|p\|_2 \|V_1\|_2,
$$
which yields \er{Y2}. From Lemma \ref{TX} we deduce that
the mapping $Y_\pm^0:\K_\pm\to \cB_1$ is analytic and satisfies
$$
\|Y_\pm^0(k)\|_{\cB_1}\le\|X_\pm(k)\|_{\cB_1}+\|T_\pm(k)\|_{\cB_1}\le {\|q-ip'\|_2 \|V_1\|_2\/\sqrt3|k|w }+{3\|p\|_2 \|V_1\|_2\/w},
$$
where $w=(|\Im k| |\Im k^+|)^{1\/2}$
which yields \er{Y1}.

From Proposition \ref{TYG} we deduce that $Y_+^0(k)-Y_-^0(k)\in \cB_1$ for all $k\ne
0$. Above we have obtained that $Y_\pm^0(k)\in \cB_1$ for all $k\ne \K_\pm$. This
yields that $Y_\pm^0(k)\in \cB_1$ for all $k\in \K_0=\K_+\cup \K_-$.
Moreover, using  $Y_+^0(k)-Y_-^0(e_-k)\in \cB_1$ for all $k\ne 0$ we
obtain $Y_\pm^0(k)\in \cB_1$ for all $k\in \K_0\cup e_+\K_0$.
Repeating these arguments give that $Y_\pm^0(k)\in \cB_1$ for all $k\in \K_0\cup e_+\K_0\cup e_-\K_0 $.

For each $k\in e_+\R\cup e_-\R, k\ne 0$ we have
$$
Y_\pm^0(k)={1\/2\pi i}\int_{|z|=\ve}{Y_\pm^0(k+z)dz\/z}
$$
for $\ve>0$ small enough. Here due to \er{Y1} and the properties of $Y_\pm^0$
 (described above) the function $\|Y_\pm^0(k+z)\|_{\cB_1}$
integrable. Thus $Y_\pm^0(k)$ is trace class for all $k\ne 0$.
\BBox

\medskip

We introduce the operator valued function $Y_\pm$ for the perturbed resolvent
$R(k)=(H-k^3)^{-1}$ by
\[
\lb{defJJ0}
Y_\pm(k) =V_2R(k)V_1,\qqq k\in \K_\pm.
\]
This operator satisfies
\[
\lb{eYY} (I-Y_\pm^0(k))(I+Y_\pm(k))=I\qqq \forall \ k\in \K_\pm.
\]

\begin{lemma}
\label{TYpe}
Let $(p,q)\in\mH$.
Then each $Y_\pm(k)\in\cB_1,k\in\K_\pm$ and  the operator-valued
function $Y_\pm: \K_\pm\to \cB_1$ is analytic and has a meromorphic
extension from $\K_\pm$ into the whole complex plane.
Moreover, $Y_\pm(k), k\in \K_\pm$ satisfies
\[
\lb{Y2x} \|Y_\pm(k)\|_{\cB_2}\le 1, \qqq if \qqq |k|\ge r_*.
\]

\end{lemma}

{\bf Proof.}  For $k\in\K_\pm$ identity \er{eYY} gives
\[
\lb{YY1x} Y_\pm(k)(k)=Y_\pm^0(k)(I+Y_\pm^0(k))^{-1}\in\cB_1
\]
and, since $Y_\pm^0$ is analytic in $\K_1$, $Y_\pm(k)$ is also analytic.
Due to the analytic Fredholm theorem, see \cite[Th VI.14]{RS72},
the function $Y_\pm(k)$ has a meromorphic  extension into the
whole complex plane.
Moreover, the identity \er{Y1} and Lemma \ref{TY} give \er{Y2x}.
\BBox

\section{The determinant}
\setcounter{equation}{0}

\subsection{The determinant.} We discuss the determinant
 $D_\pm(k), k\in \K_\pm$.

\begin{lemma}
\lb{TD1}
Let $(p,q)\in\mH$. Then the determinant $D_+(k)=\det(I+Y_+^0(k))$ is analytic in  $\K_+$ and has  an analytic  extension
 into the whole complex plane without zero. Moreover,  the function $k^3D_\pm (k)$ is entire.
\end{lemma}

\no {\bf Proof.}  Due to Lemma~\ref{TY}~ii) each  operator $Y_\pm^0(k), k\in \K_\pm$ belongs to the class
$\cB_1$ and the operator-valued
function $Y_\pm^0(k)$ is analytic in $k\in \K_\pm$. Moreover, $Y_\pm^0(k)$ has
an analytic extension from $\K_\pm$ into the whole complex
plane without zero. Then the determinant $D_\pm(k),
k\in \K_\pm$ is analytic in $k\in \K_\pm$ and has  an analytic
extension from $k\in \K_\pm$ into the whole complex plane without
zero. The statement that $k^3D_\pm (k)$ is entire  was
proved in \cite{OY12}. \BBox

\subsection{Asymptotics of the determinant}
The estimates \er{Y1} give $\|Y_+^0(k)\|_{\cB_1}=o(1)$ as $|k| \to
\infty, k\in\K_+$. Thus we can define the branch $\log D_+(k),k\in\K_+$
for such $k$ large enough by
\[
\lb{Logas} \log D_+(k) = o(1)\qqq  \as\qqq |k| \to \infty,\qq {\rm
and} \qq \qq k\in \K_+.
\]

We need the following standard results.

\begin{lemma}
\lb{TD2x}
Let $(p,q)\in\mH$. Then
\[
\lb{TrY} \Tr Y_+^0(k)={2e_+ p_0\/3ik}+{e_-q_0\/3ik^2},\qqq \forall \
k\ne0.
\]
If $k\in\K_+, |k|\ge r_*$, where $r_*>0$  is defined by \er{Cr},
then
\[
\lb{TrY12} |\Tr Y_+^0(k)|\le {1\/4},
\]
\[
\lb{2.12}
\log D_+(k)=-\sum _{n=1}^\iy {1\/n}\Tr (-Y_+^0(k))^n, \ \ \
\]
\[
\lb{2.13}
 |\log D_+(k)+\sum _{n=1}^{N}{1\/n}\Tr (-Y_+^0(k))^n|\le
{C_*^{N+1}\/|k|^{N+1} },\ \ \ \forall  \ N\ge 1,
\]
\[
\lb{D12}
|\log D_+(k)-\Tr Y_+^0(k)|\le \log 2-{1\/2}, \ \ \
\]
\[
\lb{D12x}
|\log D_+(k)|\le 2,
\]
where the series converges absolutely and uniformly in $|k|\ge r_*$.

\end{lemma}

\no {\bf Proof.} Let $k\in \K_+$. The kernel of the free resolvent
$R_0^+(k,t)$ is analytic in $t=x-y$. We determine asymptotics of
$R_0^+(k,t)$ as $t\to 0$. The identities \er{S3} give
$$
\begin{aligned}
\label{S3i} {-i3k^2}R_0^+(k,t)=
\ca e^{itk}+e_+e^{itke_+}=1+e_++ikt(1+e_+^2)+O(t^2)=h+O(t^2)\ &  \as \ t>0\\
-e_- e^{itk e_-}=-e_--e_-^2 itk+O(t^2)=h+O(t^2)\  & \as \ t<0 \ac,\\
\end{aligned}
$$
where $h=-e_--ikte_+$  as $t\to 0$, since $1+e_++e_-=0$.  Let
$\ve_0={ie_-\/3} $.
 This yields
\[
\begin{aligned}
\label{St1} R_0^+(t,k)={e_-+ikte_+\/i3k^2}+O(t^2) =-{\ve_0\/k^2}
\big(1+ike_- t\big)+O(t^2),
\\
\pa R_0^+(t,k)=-{\ve_0 e_-\/k} +O(t).
\end{aligned}
\]
Using \er{St1} we have
$$
\Tr Y_+^0(k)=\int_0^\g (2p\pa_x+q_p) R_0^+(k,t)\big|_{x=y}dx=
 -{2\ve_0 e_- p_0\/k}-{\ve_0 \/k^2}\int_\R (q-ip')dx={2e_+ p_0\/3ik}
 +{e_-q_0\/3ik^2},
$$
since $\int_\R (q-ip')dx=q_0$. This gives \er{TrY}.  We have
$$
|\Tr Y_+^0(k)|\le {2|p_0|\/3|k|}+{|q_0|\/3|k|^2}\le {2\|p\|\sqrt
\g\/3|k|}+{\|q\| \sqrt \g\/3|k|^2}\le {C_*\/6|k|}+{C_*\/6|k|^2}\le
{C_*\/3|k|}\le  {1\/4},
$$
which yields \er{TrY12}. The estimates \er{Y2}, \er{Y2c} imply
\[
\lb{estY0n}
\begin{aligned}
 \big|\Tr(Y_\pm^0(k)^n)\big|\le \|Y_\pm^0(k)\|_{\cB_2}^n
\le{(2C_*/3)^n\/|k|^{n}},\\
\big|\Tr(Y_\pm^0(k)^n)\big|\le \|Y_\pm^0(k)\|_{\cB_2}^n \le 2^{-n},
\end{aligned}
\]
for all \ $n\ge 2$. Then the series \er{2.12} converges absolutely
and uniformly and it is well-known that the sum is equal to $\log
D_+(k)$ (see \cite[Lemma XIII.17.6]{RS78}). The estimates \er{2.13}
follow from \er{estY0n}, \er{2.12}:
$$
|\log D_+(k)+\sum _{n=1}^{N}{1\/n}\Tr (-Y_+^0(k))^n|\le
{(2C_*/3)^{N+1}\/|k|^{N+1}}\rt(1+2^{-1}+2^{-2}+... \rt)\le
{C_*^{N+1}\/|k|^{N+1}}.
$$
 Moreover, in this case we have
$$
|\log D_+(k)-\Tr Y_+^0(k)|\le \sum _{n=2}^\iy {2^{-n}\/n}=   \log 2-{1\/2}, \ \ \
$$
which  gives \er{D12} and adding \er{TrY12} we have \er{D12x}.
\BBox

We determine asymptotics of $D_+(k)$ as $|k|\to \iy$ in $\K_+$.

\begin{lemma}
\lb{TD2} Let $(p,q)\in\mH$ and let $|k|\to \iy,\qq k\in \K_+$. Then
\[
\lb{TrY2} \Tr Y_+^0(k)^2={O(1)\/k^3},
\]
 \[
\lb{aD1}
 \log D_+(k)={2e_+ p_0\/3ik}+{e_-q_0\/3ik^2}+{O(1)\/k^3},
\qqq
\]
uniformly in $\arg k\in [0,{\pi\/3}]$.

\end{lemma}

\no {\bf Proof.}   The definition \er{Y0} and the identity \er{S3}
give
$$
\begin{aligned}
\Tr Y_+^0(k)^2=J(k)+{O(1)\/k^3},\qq J(k)=\iint_{[0,\g]^2}p(x)\pa_x
R_0(x-y,k) p(y)\pa_y R_0(y-z,k)|_{x=z}dydx
\end{aligned}
$$
and using \er{S4} we obtain
\[
\lb{TrY21}
\begin{aligned}
J(k)={e_+\/9k^2}\iint_{t>0} p(x)p(y)(e^{itk}+e_-
e^{itk^+})e_+e^{-itk^-}dxdy\\
+{e_+\/9k^2}\iint_{t<0} p(x)p(y)e^{itk^-} (e^{-itk}+e_-
e^{-itk^+})e^{itk^-}dxdy
\\
={e_+\/9k^2}\iint_{\R^2} p(x)p(y)(e^{i|t|(k-k^-)}+e_-
e^{i|t|(k^+-k^-)})dxdy= {O(1)\/k^3},
\end{aligned}
\]
where we have used the integration by parts in the last identity. Similar
arguments show $\Tr (Y_+^0(k))^3=O(k^{-4})$. The last asymptotics and
the estimate \er{2.13} give
$$
\log D_+(k)=\Tr Y_+^0(k)-{1\/2}\Tr (Y_+^0(k))^2+O(k^{-3}).
$$
The asymptotics \er{TrY}, \er{TrY2} and the estimate \er{2.13} give the asymptotics \er{aD1}. \BBox

Now we are ready to prove the first theorem.

{\bf Proof  of Theorem \ref{T1}.} Let $(p,q)\in\mH$. Then by Lemma
\ref{TD1},  the determinant $D_+(k)=\det(I+Y_+^0(k))$ is analytic in
$\K_+$ and has  an analytic extension  into the domain $\C\sm
\{0\}$. Moreover,  the function $k^3D_\pm (k)$ is entire.
Asymptotics \er{asDK+} was determined in Lemma \ref{TD2}.

ii)
In order to show the equalities \er{Tr1},  we need the following simple
result. Assume that a function $f$ is analytic in $\C_+$ and continuous up to
$\R\sm \{0\}$ and  satisfies the following
condition:
\begin{equation}
  {\rm Im}\,f(\cdot+i0)\in L^1({\R}), \qqq
f(i\tau)=-\frac{Q_f+o(1)}{i\tau}\quad as \qq  \tau\to \infty.
\label{S2Condtiononf}
\end{equation}
for some $Q_f\in \R$. Then
\begin{equation}
Q_f=\frac{1}{\pi}\int_R{\rm Im}\,f(t+i0)dt. \label{S2Imfkint}
\end{equation}
Then the above arguments from i) give that
the function $D_+(\l^{1\/3}), \l^{1\/3}\in \K_+$ is analytic in $ \C_+$, continuous up to $\R\sm \{0\}$. Define a function
$$
f(\l)=ie^{i{\pi\/3}} {\log D_+(\l)\/\l^{{2\/3}}},\qqq  \l\in
\C_+.
$$
Due to \er{asDK+} this function satisfies:
\[
\Re f(\l+i0)\in L^1(\R),\qqq  \qqq f(i\t)=-{2p_0+o(1)\/3(i\t) }
\qqq as \qqq \t\to \iy.
\]
This yields \er{Tr1}.
\BBox.

\section{The identities for the determinant and S-matrix}
\setcounter{equation}{0}

\subsection{The spectral representation for $H_0$}
In this section we study scattering matrix.
Recall the Fourier transformation  $\F f(k)= \wh
f(k)={1\/\sqrt{2\pi}}\int_\R f(x)e^{-ik x}dx $ for $k\in\R$. Recall
that $ V_1(x)=\c_{[0,\g]}(x)$ and $V_2= \big(2p\pa +(q-ip')\big)$.
Introduce the functionals $\p_1(k): L^2(\R)\to\C $ and $\p_2(k):\C
\to L^2(\R)$ given by
\[
\lb{defPsi}
\begin{aligned}
& (\p_1(k)f)(x)=\F V_1 f(k)={1\/\sqrt{2\pi}}\int_\R
e^{-ikx}V_1(x)f(x)dx,
\\
&  \p_2(k)c
={1\/\sqrt{2\pi}}e^{ikx}V(x,k)c,\qqq V(x,k)=k
2p+(q-ip'),\\
\end{aligned}
\]
for $f\in L^2(\R)$ and $c\in \C$.

\begin{lemma}
\label{TPk} Let $(p,q)\in\mH$. Then  the functionals $\p_j(k), k\in
\R_+, j=1,2$ have analytic extensions into the whole complex plane
and for all $k\in\C$ satisfy
\[
\label{Pe1}
\begin{aligned}
\|\p_1(k)\|\le  {\sqrt \g\/\sqrt{2\pi}}  e^{\g (\Im k)_+},
\end{aligned}
\]
\[
\label{Pe2}
\begin{aligned}
\|\p_2(k)\|\le  {1\/\sqrt{2\pi}}
\|V(\cdot,k)\|_2e^{\g (\Im k)_-}\\
\end{aligned}
\]
\[
\lb{Pe3} \p_1(k)\p_2(k)={2kp_0+q_0\/2\pi}={V_0\/2\pi},
\]
for all $k\in\C$, where $q_0=\int_\R qdx$ and
$$
(a)_\pm ={1\/2}(|a|\pm a)\qqq  \forall \ a\in \R.
$$

\end{lemma}

 {\bf Proof.} It is clear that $\p_1(k), k\in \R_+$
has an analytic extension into the complex plane. The functional
$\p_1(k), k\in \C$ satisfies
$$
\begin{aligned}
g=(\p_1(k)f)_{L^2(\R)}=\int_0^\g {e^{-ikx}\/\sqrt{2\pi}}\ f(x)dx,
\qqq |g|^2\le \g {e^{2\g(\Im k)_+}\/2\pi} \|f\|_2^2,
\end{aligned}
$$
which yields  \er{Pe1}. The proof for  $\p_2$ is similar. From the
definition  of $\p_1, \p_2$, we obtain
$$
\p_1(k)\p_2(k)={1\/2\pi}\int_\R\big(k2p+(q-ip')\big)dx={2kp_0+q_0\/2\pi}={V_0\/2\pi}.
$$
\BBox

\subsection{The scattering matrix.}

The identity $(\F H_0\F^*\wh f)(k)=k^3\wh f(k),\ k\in \R$, implies
that $\F H_0\F^*$ is the operator of multiplication by $k^3$ in
$L^2(\R,dk)$. Recall that the operator $\cS =W_+^*W_-$ commutes with
the operator $H_0$.
 Then the operator $\F\cS\F^*$ acts in the space
$L^2(\R,dk)$ as multiplication by a scalar function $S_+(k)=\cS(k^3),k>0 $
(corresponding to the spectral parameter $k^3=\l>0$) has the form:
\[
\label{Sp}
\begin{aligned}
S_+(k)=1-c_k\cA^+(k),\qqq \forall\ k>0,\qq
\cA^+=\cA_0^+-\cA_1^+,
\\
c_k={2\pi i\/3k^2}, \qq
\cA_0^+(k)=\p_1(k)\p_2(k),\qqq \cA_1^+(k)=\p_1(k)Y_+(k+i0)\p_2(k)
\end{aligned}
\]
and a scalar
function $S_-(k)=\cS(k^3),k\in e^{i{\pi\/3}}\R_+$ (corresponding to the
spectral parameter $k^3=\l<0$) given by
\[
\label{Sm}
\begin{aligned}
S_-(k)=1-c_k\cA^-(k), \qq
\cA^-(k)=e_+\p_1(k^+)(I-Y_+(k+0))\p_2(k^+)
\end{aligned}
\]
for all $k\in e^{i{\pi\/3}}\R_+$,
see, e.g., \cite{RS79}). Here $\cA^\pm(k)$ is the modified
scattering amplitude.

\begin{lemma}
\label{TAm} Let $(p,q)\in\mH$. Then the modified scattering
amplitude $\cA^+(k)$  has a meromorphic extension from $\R_+$ into
the whole complex plane. Moreover, $\cA^+(k), k\in\ol\K_+, |k|\ge
r_*$ satisfies
\[
\lb{A0} \cA_0^+(k)={2kp_0+q_0\/2\pi},\qq k\in\C,
\]
\[
\lb{A1} \|\cA_1^+(k)\|_{\cB_2}\le {\sqrt \g\/2\pi} \|V(\cdot,k)\|_2e^{\g (\Im k)_+},
\]
\[
\lb{eS} |S_+(k)-1|\le {C_*\/6|k|}(1+e^{\g (\Im k)_+}) \le
{1\/6}(1+e^{\g (\Im k)_+}).
\]
\end{lemma}

{\bf Proof.} From the definition  of $\cA_0^+(k)$ in \er{Sp}  and
\er{Pe3} we obtain \er{A0}.

From the definition  of $\cA_1^+(k)$ in \er{Sp}, \er{defPsi} and from
estimates \er{Pe1},  \er{Pe2}, \er{Y2x}  we obtain
$$
\begin{aligned}
{2\pi \/3|k|^2}|\cA_1^+(k)|=|\p_1(k)Y_+(k+i0)\p_2(k)|\le
{\sqrt\g\/ 3|k|^2}\|2p k +(q-ip')\|_2e^{\g (\Im k)_+}
\\
\le {\sqrt\g\/ 3 |k|}(2\|p\|_2 +\|q-ip'\|_2)e^{\g (\Im k)_+}=
{C_*\/6|k|}e^{\g (\Im k)_+}\le {e^{\g (\Im k)_+}\/6}
\end{aligned}
$$
and
$$
{2\pi \/3|k|^2}|\cA_0^+(k)|={(2|p_0| |k| +|q_0|)\/3|k|^2}\le
{\sqrt\g\/3|k|}(2\|p\|_2 +\|q-ip'\|_2)= {C_*\/6|k|}\le {1\/6}.
$$

Using \er{Sp} and estimates above we obtain
$$
|S_+(k)-1|\le {C_*\/6|k|}+{C_*\/6|k|}e^{\g (\Im k)_+}\le {1+e^{\g
(\Im k)_+}\/6},
$$
which yields \er{eS}.
\BBox

We describe an analytic  extension
of the scattering matrix.

\no {\bf Proof of Theorem \ref{TS}.} i) Let $\l>0$. Then for $k\pm
i0=(\l\pm i0)^{1\/3}\in \R_+\pm i0$ we have the Birman-Krein formula
\cite{BK62}
$$
{D_-(k-i0)\/D_+(k+i0)}={D_-(k)\/D_+(k)}=\cS(k^3)=S_+(k), \qqq
k=(\l)^{1\/3}\in \R_+,
$$
which yields \er{DS1}, since the functions $D_\pm $ are analytic in $\C\sm \{0\}$.
Thus the S-matrix $S_+(k), k>0$ has an analytic extension into the
sector $\K_+$ and a meromorphic extension into the whole complex
plane $\C$. The zeros of $S_+$ coincide with the zeros of $D_-$ and
the poles of $S_+$ are precisely the zeros of $D_+$.

ii) Let $\l<0$. Then for $k=(\l)^{1\/3}\in e^{i{\pi\/3}}\R_+$ we have
the Birman-Krein formula
$$
{D_-(k_1)\/D_+(k)}=\cS(k^3)=S_-(k), \qqq \where \ k_1=e^{-i2{\pi\/3}}     k\in
e^{-i{\pi\/3}}\R_-,
$$
which yields \er{DS2}, since the functions $D_\pm $ are analytic in
$\C\sm \{0\}$.
Later on the proof is similar to the case i).\BBox


Using the Birman-Krein formula we estimate the determinant $D_-(k)$ in the
sector $\K_+$.

\begin{lemma}
\lb{TY1} Let $p,q\in\mH$, and let $k\in\K_+, |k|\ge r_*$. Then
\[
\lb{eSx} |D_-(k)|\le 2 \rt(1+{C_*\/6|k|}(1+e^{\g (\Im k)_+})\rt) \le
2+ {1\/3}(1+e^{\g (\Im k)_+}).
\]
\end{lemma}

 {\bf Proof.} Substituting estimates \er{eS}, \er{D12x}  in the identity \er{DS1}
 we obtain
 $$
|D_-(k)|=|D_+(k)S_+(k)| \le 2 \rt(1+{C_*\/6|k|}(1+e^{\g (\Im
k)_+})\rt) \le 2+ {1\/3}(1+e^{\g (\Im k)_+}).
 $$
\BBox

\subsection{Transformations of the determinant}
In Section 3 we have obtained the analytic extension of the function $D_\pm$ from
$\K_\pm$ in the into $\C\sm \{0\}$. In order to get estimates of the determinant
in the complex plane we need new identities.
Now we describe these analytic extension of the function $D_\pm$ from
$\K_\pm$ into $\C\sm \{0\}$. In fact we construct the precise
analytic continuation in each of the domains $\K_\pm', \K_\pm''$
given by (see Fig. \ref{fig1}):
\[
\lb{Kx} \K_+'=e_*\K_+,\qq \K_+''=e_+\K_+,\qq and \qqq  \K_-'=\ol
e_*\K_-,\qq \K_-''=e_-\K_-.
\]
where $e_*=e^{i{\pi\/3}}$ and $e_\pm=e^{\pm i{2\pi\/3}}$.
We rewrite \er{X} in another form and determine the first identity
\[
\lb{dP}
 Y_+^0(k)-Y_-^0(k)=P(k)=c_k\p_2(k)\p_1(k),\qqq k\ne 0, \qqq
 c_k={i2\pi\/3k^2},
\]
where $P(k)$  is a rank one operator with a kernel $P(k,x,y)$ given
\er{Xxy}.

We rewrite \er{E2} in another form and determine the second identity
\[
\lb{dPe} Y_+^0(k)-Y_-^0(k^-)=P(k^+)=c_k e_+\p_2(k^+)\p_1(k^+),\qqq k\ne
0.
\]
Thus \er{dP} and \er{dP} imply the third identity for $Y_+^0$:
\[
\lb{YY1} Y_+^0(k)=Y_+^0(k^-)+P_1(k), \qq P_1(k)=P(k^+)-P(k^-).
\]
We define
\[
\lb{P2P1} \p_j^\pm(k)=\p_j( k^\pm), \qq
\P_2=(e_+\p_2^+,-e_- \p_2^-),\qqq \P_1=\ma \p_1^+\\
\p_1^-\am  .
\]
Then we have
\[
\lb{P1} P_1=P(k^+)-P(k^-)=c_k\rt(e_+\p_2^+\p_1^+-e_-
\p_2^-\p_1^-\rt)=c_k \P_2 \P_1.
\]

\begin{lemma}
\lb{TPP12} Let $p,q\in\mH$, and let $k\in\C$. Then
\[
\lb{PP1}
\begin{aligned}
\P_1(k) \P_2(k)
={1\/2\pi}\int_0^\g \ma  e_+ V(x,k^+) & -e_- e^{\sqrt3 kx }V(x,k^-) \\
e_+e^{-\sqrt3 kx }V(x,k^+) & -e_-V(x,k^-)\am dx.
\end{aligned}
\]
\end{lemma}

 {\bf Proof.} Substituting  \er{P2P1}, \er{defPsi}  in the identity \er{DS1}
 we obtain
 $$
 \begin{aligned}
\P_1(k) \P_2(k)=\ma \p_1^+(k)\\
\p_1^-(k)\am (e_+\p_2^+(k),-e_- \p_2^-(k))
\\
={1\/2\pi}\int_0^\g
\ma e^{-ik^+x } \\
e^{-ik^-x }\am V(x) (e_+ e^{ik^+x },-e_- e^{ik^-x })dx
={1\/2\pi}\ma e_+ V_0(k^+) & 0\\
0 &  -e_- V_0(k^-) \am+\P_{12}(k),
\end{aligned}
 $$
and using $-i(k^+-k^-)=\sqrt 3 k$ we have
 $$
 \begin{aligned}
\P_{12}(k)={1\/2\pi}\int_0^\g \ma  0& -e_- e^{-i(k^+-k^-)x }V(x,k^-) \\
e_+e^{i(k^+-k^-)x }  V(x,k^+) & 0\am dx\\
={1\/2\pi}\int_0^\g \ma  0& -e_- e^{\sqrt3 kx }V(x,k^-) \\
e_+e^{-\sqrt3 kx } V(x,k^+) & 0\am dx.
\end{aligned}
 $$
\BBox

\begin{lemma}
\lb{TPPP} Let $p,q\in\mH$, and let $k\ne 0$. Then
\[
\lb{PPP1} P(k^+)-P(k^-)+P(k) =c_k \big(\P_2(k),\p_2(k)\big) \ma \P_1(k)\\
\p_1(k)\am ,
\]
\[
\lb{PPP2}
\begin{aligned}
\ma \P_1\\
\p_1\am(\P_2,\p_2) =\ma \P_1 \P_2 & \P_1 \p_2\\
\p_1 \P_2 & \p_1 \p_2 \am  ,
\end{aligned}
\]
where
\[
\lb{PPP3}
\begin{aligned}
\p_1(k) \P_2(k)={1\/2\pi}\int_0^\g (e_+ V(x,k^+) e^{-\sqrt 3 e_*kx
},-e_- V(x,k^-)e^{\sqrt 3\ol e_* kx })dx,
\\
\P_1(k) \p_2(k)= {1\/2\pi}\int_0^\g \ma e^{\sqrt 3 e_*kx } \\
e^{-\sqrt 3 \ol e_*kx }  \am V(x,k) dx,
\end{aligned}
\]
and $c_k={i2\pi\/3k^2}$ and $e_*=e^{i{\pi\/3}}$.

\end{lemma}

 {\bf Proof.} Substituting  \er{P2P1}, \er{defPsi}  in the identity \er{DS1}
 we obtain
  $$
\ma \P_1\\
\p_1\am(\P_2,\p_2)  =\ma \P_1 \P_2 & \P_1 \p_2\\
\p_1 \P_2 & \p_1 \p_2 \am .
$$
Thus we need to consider the terms $\P_1 \p_2 $ and $\p_1 \P_2$,
since the term $\P_1 \P_2$ and $\p_1 \p_2$ has been studied in
Lemmas \ref{TPP12} and \ref{TAm}. Substituting  \er{P2P1},
\er{defPsi} in the identity \er{DS1}
 we obtain
  $$
 \begin{aligned}
\p_1(k) \P_2(k)=\p_1(k) (e_+\p_2^+(k),-e_- \p_2^-(k))
={1\/2\pi}\int_0^\g e^{-ikx } V(x,\pa) (e_+ e^{ik^+x },-e_- e^{ik^-x
})dx\\
={1\/2\pi}\int_0^\g (e_+ V(x,k^+) e^{ik(e_+-1)x },-e_-
V(x,k^-)e^{ik(e_--1)x })dx
\end{aligned}
 $$
where $i(e_+-1)=-\sqrt 3 e^{i{\pi\/3}}$ and  $i(e_--1)=\sqrt 3
e^{-i{\pi\/3}}$. Similar arguments imply
 $$
 \begin{aligned}
\P_1(k) \p_2(k)= \ma \p_1^+(k)\\
\p_1^-(k)\am  \p_2(k)={1\/2\pi}\int_0^\g \ma e^{-i(k^+-k)x } \\
e^{-i(k^--k)x }  \am V(x,k) dx.
\end{aligned}
 $$

\BBox

In order to prove the main estimate \er{esDK} we need the following identities.

\begin{lemma}
\lb{TDK} Let $p,q\in\mH$, and let $k\ne 0$. Then the following
identities hold true:
\[
\lb{DK1} D_+(k) =D_-(k^-)\det (I+c_k\p_1(k^+)\cJ_-(k^-)\p_2(k^+)),
 \qqq k\in \K_+',
\]
\[
\lb{DK2}
\begin{aligned}
D_+(k)=D_+(k^-) \det (I+c_k \P_1(k) \cJ_+(k^-)\P_2(k)  ) ,\qqq k\in
\K_+'',
\end{aligned}
\]
and
\[
\lb{DK3z}
\begin{aligned}
D_+(k)=D_-(k^+) \det \rt(I+c_k \ma \P_1(k)\\ \p_1(k)\am \cJ_-(e_+k)
(\P_2(k),\p_2(k)) \rt), \qqq k\in \K_-'',
\end{aligned}
\]
where $\cJ_\pm(k)=I-Y_\pm(k)$.
\end{lemma}

 {\bf Proof.}
1) Let $k\in \K_+'$.  Then $k^-\in \K_-$ and due to \er{dPe} we have:
$$
\begin{aligned}
D_+(k)=\det (I+Y_+^0(k))=\det (I+Y_-^0(k^-)+P(k^+))=
D_-(k^-)\det (I+\cJ_-(k^-)P(k^+)) \\
\end{aligned}
$$
which  gives \er{DK1}.

2)  Let $k\in \K_+''$.   Then $k^-\in \K_+$ and  due to \er{YY1} we
have:
$$
\begin{aligned}
D_+(k)=\det (I+Y_+^0(k))=\det (I+Y_+^0(k^-)+P_1(k)) \\
=\det (I+Y_+^0(k^-))\det (I+\cJ_+(k^-)P_1(k))=D_+(k^-) \det
(I+\cJ_+(k^-)c_k \P_2(k) \P_1(k))\\
=D_+(k^-) \det (I+c_k \P_1(k) \cJ_+(k^-) \P_2(k) ).
\end{aligned}
$$

3)   Let $k\in \K_-''$, where $k^+\in\K_-$. We have
\[
\begin{aligned}
Y_+^0(k)=Y_+^0(k^+)-P_1(k^+)=Y_-^0(k^+)+P(k^+)-P(k^-)+P(k)\\
=Y_-^0(k^+)+c_k (\P_2(k),\p_2(k)) \ma \P_1(k)\\ \p_1(k)\am .
\end{aligned}
\]
Then we obtain
$$
\begin{aligned}
D_+(k)=\det (I+Y_+^0(k))=\det \rt(I+Y_-^0(k^+)+P(k^+)-P(k^-)+P(k)\rt) \\
=\det(I+Y_-^0(k^+)) \det \rt(I+c_k \cJ_-(k^+) (\P_2(k),\p_2(k)) \ma
\P_1(k)\\ \p_1(k)\am\rt)\\
=D_-(k^+) \det \rt(I+c_k \ma \P_1(k)\\ \p_1(k)\am \cJ_-(k^+)
(\P_2(k),\p_2(k)) \rt).
\end{aligned}
$$ which  gives \er{DK3z}.
 \BBox

\section{Estimates of the determinant on the plane }
\setcounter{equation}{0}

 We prove the basic estimates \er{esDK} of
$D_+$ in the plane.


\begin{lemma}
\lb{TaD1} Let $p,q\in\mH$, and let $|k|\ge r_*=\max \{{4\/3}C_*,1\}$, where
$C_* =2\sqrt\g \big(\|q-ip'\|_2+2\|p\|_2\big)$. Then
\[
\lb{DK1e} |D_+(k)|\le 2+ e^{\g|\Im k^+|},\qqq k\in \K_+',
\]
\[
\lb{DK2e}
\begin{aligned}
|D_+(k)|\le 4 e^{-\g \sqrt3 \Re k}, \qqq k\in \K_+''.
\end{aligned}
\]
\end{lemma}

 {\bf Proof.}
1) Let $k\in \K_+'$.  Then we have $k^+\in \K_-'', k^-\in \K_-$. Due to
\er{dPe} the determinant $D_+(k)$ has the form
\[
\lb{ED1}
\begin{aligned}
D_+(k) =D_-(k^-)\det (I+c_kF(k)),\qq
 F=F_0-F_1,\\ F_1(k)=
\p_1(k^+)Y_-(k^-)\p_2(k^+),\qqq F_0(k)=\p_1(k^+)\p_2(k^+),
\end{aligned}
\]
where $c_k={i2\pi\/3k^2}$.
Lemma \ref{TPk} gives
\[
\lb{F0x}
F_0(k)=\p_1(k^+)\p_2(k^+)={V_0(k^+)\/2\pi}.
\]
The resolvent estimate \er{Y2x} and  Lemma \ref{TPk} gives
\[
\lb{F0xx}
\begin{aligned}
|F_1(k)|\le {\sqrt \g\/2\pi}\|V(\cdot, k^+)\|_2\|Y_-(k^-)\|
e^{\g|\Im k^+|}\le {\sqrt \g\/2\pi} v(k^+) e^{\g|\Im
k^+|},
\\ \where \ v(k)=\|V(\cdot, k)\|_2.
\end{aligned}
\]
Combine \er{F0x}-\er{F0xx} and \er{V1}  we obtain
$$
|c_k||F(k)|\le {|c_k||V_0(k^+)|\/2\pi}+{|c_k|\sqrt
\g\/2\pi}\|V(\cdot, k^+)\|_2 e^{\g|\Im k^+|}\le
|c_k|{|k|C_*\/\pi}e^{\g|\Im k^+|}\le {e^{\g|\Im k^+|}\/2}
$$
since ${C_*\/3|k|}\le {1\/4}$. Then due to \er{D12x}, $|D_-(k^-)|\le 2$
and \er{ED1} we have \er{DK2e}.

2)  Let $k\in \K_+''$.  Then we have $k^+\in \K_-'$ and $k^-\in \K_+$.  Due to
\er{DK2} the determinant $D_+(k)$ has the form
$$
\begin{aligned}
D_+(k)=D_+(k^-) \det (I+c_k F(k)  ),\qqq F=F_0-F_1,\\
F_0(k)=\P_1(k) \P_2(k),\qqq F_1(k)= \P_1(k)Y_+(k^-)\P_2(k),
\end{aligned}
$$
Lemma \ref{TPP12}
gives
\[
\lb{PP1a}
\begin{aligned}
F_0(k)=\P_1(k) \P_2(k)
={1\/2\pi}\int_0^\g \ma  e_+ V(x,k^+)& -e_- e^{\sqrt3 kx }V(x,k^-) \\
e_+e^{-\sqrt3 kx }V(x,k^+) & -e_-V(x,k^-)\am dx,
\end{aligned}
\]
and then
\[
\lb{PP1b}
\begin{aligned}
|F_0(k)|_M\le_\bu {1\/2\pi} \ma  |V_0(k^+)|& \sqrt\g v(k^+) \\
\sqrt\g e^{-\g \sqrt3\Re k} v(k^+) & |V_0(k^-)|\am
\le_\bu {|k|C_*\/2\pi} \ma  1& 1 \\
 e^{-\g \sqrt3\Re k}  & 1\am  .
\end{aligned}
\]
Here and below for the matrix $b=\{b_{ij}\}$ we define $|b|_{M} =\{|b_{ij}|\}$
and we will write:
\[
\lb{M}
\begin{aligned}
&  |b|_{M}:=\{|b_{ij}|\},\\
&   \{b_{ij}\}\le_\bu \{c_{ij}\} \ \Lra \qq b_{ij}\le c_{ij}
\qq \forall \ ij.
\end{aligned}
\]

Definition \er{P2P1} gives
\[
\begin{aligned}
F_1= \P_1Y_+(k^-)\P_2=\ma \p_1^+\\
\p_1^-\am  Y_+(k^-)(e_+\p_2^+,-e_- \p_2^-)
\\
=\ma \p_1^+ Y_+(k^-)e_+\p_2^+  & -e_-\p_1^+ Y_+(k^-) \p_2^-  \\
\p_1^-Y_+(k^-)e_+\p_2^+ &   -e_-\p_1^-Y_+(k^-) \p_2^-\am
\end{aligned}
\]
Then due to \er{V1}, \er{V1} we estimate the matrix $F_1(k)$ by
\[
\begin{aligned}
|F_1(k)|_M \le_\bu
\|Y_+(k^-)\| {\sqrt\g\/2\pi}\ma e^{\g|\Im k^+|}  v(k^+)  & v(k^-)  \\
e^{\g(\Im k^-)_++\g(\Im k^+)_-}v(k^+) &   e^{\g|\Im
k^-|} v(k^-)\am
\\
\le_\bu
 {\sqrt\g\/2\pi}\ma e^{\g|\Im k^+|} v(k^+)  & v(k^-)  \\
e^{-\g \sqrt3 \Re k} v(k^+) & e^{\g|\Im k^-|}v(k^-)\am \le_\bu
 {C_* |k|\/2\pi}\ma e^{\g|\Im k^+|}  & 1  \\
e^{-\g \sqrt3 \Re k} & e^{\g|\Im k^-|}\am.
\end{aligned}
\]
and then
$$
\begin{aligned}
|D_+(k)|\le |D_+(k^-)| | \det (I+c_k F(k))|\le 2 \det G \le 4 e^{-\g
\sqrt3 \Re k},
\\
G={1\/16}\det \ma 5+e^{\g|\Im k^+|}  & -1  \\
e^{-\g \sqrt3 \Re k} & 5+e^{\g|\Im k^-|}\am,
\end{aligned}
$$

since ${C_*\/3|k|}\le {1\/4}$ and $|\Im k^+|+|\Im k^-|=-\sqrt3 \Re
k$. \BBox

We estimate $D_+$ in the most difficult case,
when $k\in \K_-''$ and $|k|\ge r_*$.

\begin{lemma}
\lb{TaD2} Let $p,q\in\mH$. Let $|k|\ge r_*$ and $\f=\arg k\in [\pi,
\pi+{\pi\/6}]$. Then
\[
\lb{DK3}
\begin{aligned}
|D_+(k)|\le 48e^{-2r\g \sin ({\pi\/3}+\f)}.
\end{aligned}
\]
\end{lemma}

 {\bf Proof.}  Let $k=re^{i\f}, \f\in [\pi, \pi+{\pi\/6}]$.  Then $k^+\in \K_-, k^-\in \K_+'$.
We need
\[
\begin{aligned}
\textstyle
\Im k^-=-r\sin (\f+{\pi\/3}),\qqq \Im k^+=-r\sin (\f-{\pi\/3}),
\\
\textstyle (\Im k^-)_++(\Im k^+)_-=-r\sin (\f+{\pi\/3})+r\sin
(\f-{\pi\/3})=-r\sqrt3 \cos \f=-\sqrt3\Re k,
\\
\textstyle (\Im k^-)_++(\Im k)_-=-r\sin (\f+{\pi\/3})+r\sin \f=-r
\cos (\f+{\pi\/6})=-\Re k e^{i{\pi\/6}}.
\end{aligned}
\]
Due to \er{DK3z} the determinant $D_+(k)$ has the form
 $$
D_+(k)=D_-(k^+) \det \rt(I+c_k F(k) \rt), \qqq F= \ma \P_1\\
\p_1\am (I-Y_-(e_+\cdot)) (\P_2,\p_2)=F_0-F_1,
 $$
 where $c_k={i2\pi\/3k^2}$ and
$$
\begin{aligned}
F_0=\ma \P_1\\ \p_1\am  (\P_2,\p_2)=\ma \P_1 \P_2 & \P_1 \p_2\\
\p_1 \P_2 & \p_1 \p_2 \am,\qqq F_1=\ma \P_1\\
\p_1\am Y_-(e_+\cdot) (\P_2,\p_2).
\end{aligned}
$$
From Lemmas \ref{TPP12} and \ref{TPPP} we have
$$
\begin{aligned}
F_0(k)= {1\/2\pi}\int_0^\g \ma  e_+ V(x,k^+) & -e_- e^{\sqrt3 kx
}V(x,k^-) & e^{\sqrt 3 e_*kx }V(x,k)
\\
e_+e^{-\sqrt3 kx }V(x,k^+) & -e_-V(x,k^-) & e^{-\sqrt 3 \ol e_*kx }V(x,k)
\\
 e_+ e^{-\sqrt 3
e_*kx } V(x,k^+) &  -e_- e^{\sqrt 3\ol e_* kx } V(x,k^-) & V(x,k)
\am dx,
\end{aligned}
$$
where $e_*=e^{i{\pi\/3}}$,  and then (here we use definitions \er{M} and
$v(k)=\|V(\cdot, k)\|_2$ )
$$
\begin{aligned}
|F_0(k)|_M\le_\bu {\sqrt \g\/2\pi} \ma  v(k^+) &
v(k^-) & v(k)
\\
e^{-\sqrt3 \g \Re k }v(k^+) & v(k^-) &
e^{-\sqrt 3 \g \Re \ol e_*k } v(k)
\\
 e^{-\sqrt 3 \g \Re
e_*k } v(k^+) &  v(k^-) & v(k)
\am
  \end{aligned}
$$
$$
\begin{aligned}
\le_\bu {C_*|k|\/2\pi} \ma  1 & 1 & 1
\\
e^{-\sqrt3 \g \Re k } & 1 & e^{-\sqrt 3 \g \Re \ol e_*k }
\\
 e^{-\sqrt 3 \g \Re
e_*k }  &  1 & 1 \am
\le_\bu
 {|k|^2\/8\pi} \ma  1 & 1 & 1
\\
e^{-\sqrt3 \g \Re k } & 1 & e^{-\sqrt 3 \g\Re \ol e_*k }
\\
 e^{-\sqrt 3 \g \Re
e_*k }  &  1 & 1 \am .
\end{aligned}
$$

Consider $F_1$. The definitions \er{P2P1} give
\[
\begin{aligned}
F_1=\ma \P_1\\
\p_1\am Y_-(e_+\cdot) (\P_2,\p_2)
\\
=\ma \p_1^+ Y_+(e_+\cdot)e_+\p_2^+  & -e_-\p_1^+ Y_+(e_+\cdot)
\p_2^- & \p_1^+Y_+(e_+\cdot)\p_2
\\
\p_1^-Y_+(e_+\cdot)e_+\p_2^+ &   -e_-\p_1^-Y_+(e_+\cdot) \p_2^-
& \p_1^-Y_+(e_+\cdot)\p_2
\\
\p_1Y_+(e_+\cdot)e_+\p_2^+ & -e_-\p_1Y_+(e_+\cdot)\p_2^-     &
\p_1Y_+(e_+\cdot)\p_2 \am .
\end{aligned}
\]
Then  Lemma \ref{TPk} and the estimate \er{Y2x} give
$$
\begin{aligned}
|F_1(k)|_M\le_\bu  {C_*|k|\/2\pi}  \ma e^{\g|\Im k^+|}  & e^{\g(\Im
k^+)_++\g(\Im k^-)_-}& e^{\g(\Im k^+)_++\g(\Im k)_-}
\\
e^{\g(\Im k^-)_++\g(\Im k^+)_-} &   e^{\g|\Im k^-|}  & e^{\g(\Im
k^-)_++\g(\Im k)_-}
\\
e^{\g(\Im k)_++\g(\Im k^+)_-}  & e^{\g(\Im k)_++\g(\Im k^-)_-}&
e^{\g|\Im k|}
 \am
\end{aligned}
$$
and then
$$
\begin{aligned}
|F_1(k)|_M\le_\bu  {|k|^2\/8\pi}\ma e^{\g|\Im k^+|}  & 1  & e^{\g(\Im
k)_-}
\\
e^{ -\g\sqrt3\Re k} &   e^{\g|\Im k^-|}  & e^{  -\g\Re k
e^{i{\pi\/6}}}
\\
e^{\g(\Im k^+)_-}  & 1  & e^{\g|\Im k|}
 \am
 .
\end{aligned}
$$
Thus
$$
\begin{aligned}
|c_k F|_M\le_\bu {1\/6} \ma  1 & 1 & 1
\\
e^{-\sqrt3 \g \Re k } & 1 & e^{-\sqrt 3 \g\Re \ol e_*k }
\\
 e^{-\sqrt 3 \g \Re
e_*k }  &  1 & 1 \am
+
{1\/6} \ma e^{\g|\Im k^+|}  & 1  & e^{\g(\Im k)_-}
\\
e^{ -\g\sqrt3\Re k} &   e^{\g|\Im k^-|}  & e^{  -\g\Re k
e^{i{\pi\/6}}}
\\
e^{\g(\Im k^+)_-}  & 1  & e^{\g|\Im k|}
 \am
  \end{aligned}
$$
$$
\begin{aligned}
 \le
 {1\/3}
\ma e^{\g|\Im k^+|}  & 1  & e^{\g(\Im k)_-}
\\
e^{ -\g\sqrt3\Re k} &   e^{\g|\Im k^-|}  & e^{  -\g\Re k
e^{i{\pi\/6}}}
\\
e^{\g(\Im k^+)_-}  & 1  & e^{\g|\Im k|}
 \am
 .
 \end{aligned}
$$
These estimates imply
\[
\lb{esF}
\begin{aligned}
 \textstyle
 |\det (I+c_k F)|\le 8(4/3)^3e^{\g(|\Im k^+|+|\Im k^-|+|\Im k|)}.
\end{aligned}
\]
Here we have for $k=re^{i\f}$:
\[
\begin{aligned}
\textstyle
|\Im k^-|+|\Im k^+|+|\Im k|=-2r({\sqrt3\/2} \cos \f+{1\/2}\cos \f)=
-2r \sin ({\pi\/3}+\f).
\end{aligned}
\]
Substituting estimates \er{esF} and \er{D12x} into the identity
$D_+(k)=D_-(k^+) \det \big(I+c_k F(k) \big)$ we obtain \er{DK3}. \BBox

\section{Proof of main theorems}
\setcounter{equation}{0}

\subsection{Asymptotics of the counting function}

 {\bf Proof of Corollary \ref{TN}.}
 Define the entire function $F(k)=D_+(k)k^m$.
 Let $\cN(r)$ be the number of zeros of an analytic function $F$ in the
disc $|k|<r$ counted with multiplicity. Recall the Jensen's formula
for $F$:
\[
\lb{Jen} \int_0^r\fr{\cN(t,F)}{t}dt= \fr{1}{2\pi}\int_0^{2\pi}\log
|F(re^{i\vt})|d\vt -\log|F(0)|.
\]

 Let $\cN(r)=\cN(r,F)$. Let $
f(re^{i\vt})=\log |D_+(re^{i\vt})|$. We take  Jensen's formula \er{Jen} at $2r$ and $r\ge r_*$
and take a difference we obtain
\[
\lb{J1}
\begin{aligned}
\int_{r}^{2r}\fr{\cN(t)}{t}dt= \fr{1}{2\pi}\int_0^{2\pi} \rt( \log
|F(2re^{i\vt})|-\log |F(re^{i\vt})|\rt)d\vt
\\
=\fr{1}{2\pi}\int_{\T}(\log |D_+(2re^{i\vt})|-\log |D_+(re^{i\vt})|)d\vt+m\log
2
\end{aligned}
\]

Then substituting inequalities $|D_+(k)|\le 48e^{2\g r}$ for all $|k|\ge r_*$ from Theorem \ref{TN}  into the
last integral we get
\[
\lb{J+}
\begin{aligned}
\cN(r) \log 2\le \int_{2}^{2r}\fr{\cN(t)}{t}dt
\le 2\log 48+6\g r+m\log 2,
\end{aligned}
\]
which yields \er{TN}.\BBox

{\bf Proof of Corollary~\ref{Ttr}}. We discuss the determinant
$D_+(k), k\in\K_+$, when the coefficients $(p,q)\in \mH$.  In
this case due to Proposition \ref{TGH} the operator $R_0(k)-R(k)\in \cB_1$, which yields the identity
\[
\lb{D1} {1\/3k^2}{D_+'(k)\/D_\pm(k)}=\Tr R_0(k)VR(k)=\Tr
(R_0(k)-R(k)), \qqq k \in \K_+.
\]
Recall that this fact is well-known for large class of operators.

We need also the well known result about the Hadamard factorization.
The function  $k^mD_+(k)$ is entire, of exponential type,  and denote by $(k_n)_{n=1}^{\iy}
$ the sequence of its zeros $\neq 0$ (counted with multiplicity), so
arranged that $0<|k_1|\leq |k_2|\leq \dots$. Then we have
\[
\lb{HFax} D_+(k)={C\/k^m}e^{bk}\lim_{r\to \iy}\prod_{|k_n|\leq
r}\lt(1-{k\/k_n}\rt)e^{{k\/k_n}},\ \ \ \ C=\lim_{k\to 0}k^mD_+(k)\ne
0,
\]
for some $b\in \C$ and  integer $m$, where the product converges uniformly in every
bounded disc. This gives
\[
\lb{DdD}
{D_+'(k)\/D_+(k)}= -{m\/k}+b+\sum_{n\ge 1}{k\/ k_n(k-k_n)},\qqq k\in
\K_+,
\]
where the sum converges uniformly on compact subsets of $\C\sm \{0, k_n, n\ge 1\}$.
Collecting all these identities we obtain \er{RR0}.

The identity \er{DS1} gives the standard formula
$$
S_+(k)={\ol D_+(k)\/D_+(k)}=e^{-2i\f_{sc}(k)}, k>0,
$$
where $\f_{sc}$ is defined by \er{dfsc}.
Differentiating this identity we obtain
$$
S_+(k)'=-2i\f_{sc}'(k)S_+(k)=\rt({\ol D_+'(k)\/\ol D_+(k)}  - {D_+'(k)\/D_+(k)}\rt)S_+(k)
$$
and substituting \er{DdD} we obtain \er{BW}.
\BBox


\section{Appendix}
\setcounter{equation}{0}

In this section we consider the scattering amplitude and
for simplicity we assume that $p=0$.
We consider the "approximation" of the scattering amplitude.
For three order operators it is not simple problem.
We show that
the second term $\cA_1^+(k)$ in \er{scA} still is not a "good" approximation.

Recall that due to \er{Sp}, \er{A0} we have that the S-matrix for $k>0$ has the form
\[
\lb{scA}
\begin{aligned}
S_+(k)=1-{2\pi
i\/3k^2}\rt( {q_0\/2\pi}-  \cA_1^+(k)\rt), \qq
\cA_1^+(k)=\p_1(k)Y_+(k+i0)\p_2(k),\qq   k>0.
\end{aligned}
\]
Here the coefficient ${q_0\/2\pi}$ corresponds to the so-called Born term.
It is a first approximation of the scattering amplitude.

Consider the next term $\cA_1^+$.
 Using the identity $Y_+=Y_+^0-Y_+^0Y_+$
we rewrite $\cA_1^+$ in the form
\[
\lb{AA1}
\begin{aligned}
\cA_1^+=T-\wt T,\qq
T=\p_1(k)Y_+^0(k+i0)\p_2(k),\qqq \\
\wt T=\p_1(k)Y_+^0(k+i0)Y_+(k+i0)\p_2(k).
\end{aligned}
\]
Here $T$ is the second approximation of $\cA_1^+$ and $\wt T$
is the third approximation.

Consider the second  term $T$. Let $t=x-y$. From \er{AA1}, \er{defPsi} we obtain
the decomposition
$$
\begin{aligned}
T_{}(k)={1\/2\pi}\iint_{\R^2} e^{-ik(x-y)} q(x)
R_+^0(k+i0,t)q(y)dxdy
=T_1(k)+T_2(k),\\
T_1(k)={1\/2\pi}\iint_{t>0}e^{-ikt} q(x)
R_+^0(k+i0,t)q(y)dxdy.
\end{aligned}
$$
$ \bu $  Consider the function  $T_1$. Then from \er{S3} we get
$$
\begin{aligned}
T_1(k) ={i\/6\pi k^2}\iint_{t>0}e^{-ikt}
q(x)q(y)(e^{itk}+e_+e^{itke_+})dxdy
\\
=\o \iint_{t>0}q(x)q(y)(1+e_+e^{it\z })dxdy=\o \rt(f_+(0) +e_+
f_+(\z)\rt),\qq
\z=(e_+-1)k,
\end{aligned}
$$
where $\o={i\/6\pi k^2}$ and $f_+(0)$ is a constant and
\[
\lb{zT0}
\begin{aligned}
f_{\pm}(\z)=\iint_{\pm t>0}q(x)q(y)e^{it\z}dxdy,\qq
 \\
e_+-1=\sqrt 3e^{i\pi -i{\pi\/6}},\qqq
  \z=(e_+-1)k\in \gS=\{\arg z \in
[\pi-{\pi\/6},\pi+{\pi\/6} ]\}.
\end{aligned}
\]
We consider  large $\z$ and let $|\z|\to \iy$. Here we have two cases

firstly, if $\z\in \gS
\cap\C_-$, then $|f_+(\z)|\to +\iy$;

secondly, if $\z\in \gS
\cap\C_+$, then $f_+(\z)=o(1)$.

$ \bu $   Consider the function $T_2$. We have
$$
\begin{aligned}
T_2(k)={1\/2\pi}\iint_{t<0}e^{-ikt} q(x) R_+^0(k+i0,t)q(y)dxdy=\\
=-e_-\o\iint_{t<0}q(x)q(y)e^{itk (e_--1)}dxdy
=-e_-\o\iint_{t<0}q(x)q(y)e^{it\z e_*}dxdy=-e_-\o f_-(e_*\z ),\\
\end{aligned}
$$
where $e_*=e^{i{\pi\/3}}$ and $e_*\z \in \C_-$. Thus we deuce that
the function $f_-(e_*\z )\to 0$ as $|k|\to \iy$.
Thus combine these two cases we obtain
\[
\lb{x2}
\begin{aligned}
T(k)=\o(f_+(0) +e_+ f_+(\z)-e_-f_-(e_*\z )),\\
f_-(e_*\z )=o(1),\qqq
f_+(\z)\to \ca 0 & as \qq \arg k\in [0,\pi/6]\\
\iy  & as \qq k\in \arg k\in [\pi/6, \pi/3]\ac
\end{aligned}
\]
as $k\in \K_+$ and $|k|\to \iy$. Then we deduce that $T(k)=\o (f_+(0)+o(1))$
as $|k|\to \iy$ and $\arg k\in [0,\pi/6]$, i.e., it does not go to $\iy$. Thus
S-matrix   for the operator $H$ has more complicated structure than for Schr\"odinger operators, since the first term $\cA_0$ and even the second term $T$ do not give information about the resonances.

\

\no\small {\bf Acknowledgments.}
Our  study was supported by the RSF grant  No.
15-11-30007.


\begin{thebibliography}{9999}
\setlength{\itemsep}{-\parskip}
\footnotesize

\bibitem[A99]{A99} L. Amour, Determination of a third-order
operator from two of its spectra,
SIAM J. Math. Anal. 30 (1999) 1010–-1028.

\bibitem[A01]{A01} L. Amour, Isospectral flows of third order operators,
SIAM J. Math. Anal. 32 (2001), 1375–-1389.

\bibitem[BK14]{BK14} Badanin, A.; Korotyaev, E. Third order operator with
periodic coefficients on the real line,  St. Petersburg Mathematical Journal, 25.5 (2014), 713--734.

\bibitem[BK12]{BK12}
Badanin, A.; Korotyaev, E. Spectral asymptotics for the
third order operator with periodic coefficients, J. Differential
Equations, 253 (2012), No 11, 3113--3146.

\bibitem[B85]{B85}
Beals R. The inverse problem for ordinary differential
operators on the line American Journal of Mathematics. 107(1985), 281--366.

\bibitem[BDT88]{BDT88} R. Beals, P. Deift, C. Tomei,
Direct and inverse scattering on the line,
Mathematical surveys and monograph series, No. 28, AMS, Providence, 1988.

\bibitem[BK62]{BK62} M. Sh. Birman, M.; Krein, M. On the theory of wave operators and scattering operators,
Dokl. Akad. Nauk SSSR, Ser. Mat., 144 (1962), 475–-478; English transl.: Soviet Math. Dokl., 3 (1962), 740–-744.


\bibitem[BZ02]{BZ02} L.V. Bogdanov, V.E. Zakharov, The Boussinesq equation
revisited, Physica D: Nonlinear Phenomena 165:3-4 (2002) 137--162.

 \bibitem[BKW03] {BKW03}  Brown, B.; Knowles, I.; Weikard, R. On the
inverse resonance problem, J. London Math. Soc. (2) 68 (2003), no. 2, 383--401.


\bibitem[C80]{C80}
Caudrey, P. J.  The inverse problem for the third order equation $u'''+ qu'+ru=- i\z^3 u$. Physics Letters A, 79(4) (1980), 264--268.

\bibitem[C06] {C06} Christiansen, T. Resonances for steplike potentials: forward and inverse results. Trans. Amer. Math. Soc. 358 (2006), no. 5, 2071--2089.



\bibitem[DTT82]{DTT82} P. Deift, C. Tomei, E. Trubowitz,
Inverse scattering and the Boussinesq equation,
Comm. on Pure and Appl. Math. 35 (1982) 567--628.

\bibitem [F84]{F84}  Firsova, N. Resonances of the perturbed Hill
operator with exponentially decreasing extrinsic potential.
Mathematical Notes, 36(1984), no 5, 854–-861.


\bibitem  [F97]{F97} Froese, R. Asymptotic distribution of resonances in one
dimension. J. Diff. Eq. 137 (1997), no. 2, 251--272.

\bibitem[H99] {H99} Hitrik, M. Bounds on scattering poles in one dimension.
Comm. Math. Phys. 208 (1999), no. 2, 381--411.



\bibitem  [IK14]{IK14}   Iantchenko, A.; Korotyaev, E.  Resonances for 1D massless
Dirac operators, Journal of Differential Equation, 256(2014),
No 8,  3038 - 3066.


\bibitem  [Ko88] {Ko88} Koosis, P.
The logarithmic integral I, Cambridge Univ. Press, Cambridge,
London, New York 1988.


\bibitem   [K04]{K04}   Korotyaev, E. Inverse resonance scattering on
the half line.  Asymptot. Anal. 37 (2004), no. 3-4, 215--226.

\bibitem  [Ko04] {Ko04}  Korotyaev, E. Stability for inverse resonance problem. Int. Math. Res. Not. 2004, no. 73, 3927--3936.


\bibitem   [K05]{K05}   Korotyaev, E. Inverse resonance scattering on
the real line. Inverse Problems 21(2005),  no.  1,  325--341.

\bibitem   [K11]{K11} Korotyaev, E. Resonance theory for perturbed Hill
operator, Asymp. Anal. 74(2011), no. 3-4, 199--227.

\bibitem   [KS12]{KS12} Korotyaev, E.;  Schmidt, K. On the resonances and
eigenvalues for a 1D half-crystal with localized impurity,  J. Reine
Angew. Math.  2012, Issue 670, 217--248.

\bibitem[K16]{K16}
 Korotyaev, E. Estimates of 1D resonances in terms of
potentials, to be published in Journal d'Analyse Mathematique.

\bibitem[Ko16]{Ko16}
 Korotyaev, E. Resonances  for 1d Stark operators, to be
published in Journal of Spectral Theory.


\bibitem[LL65]{LL65} Landau L. D., Lifshits, E. M. Quantum Mechanics, Non-relaticistic Theory. – Perganon Press, 1965.

\bibitem[L66]{L66}
Leibenzon, Z. I. The inverse problem of the spectral analysis of
ordinary differential operators of higher orders. Trudy Moskovskogo
Matematicheskogo Obshchestva 15 (1966), 70--144.

\bibitem   [MSW10]{MSW10}
 Marletta, M.;  Shterenberg, R.; Weikard, R.,
On the Inverse Resonance Problem for Schr\"odinger Operators,
Commun. Math. Phys., 295(2010), 465--484.



\bibitem[Ne07]{Ne07} Nedelec, N. Asymptotics of resonances for a
Schr\"odinger operator with vector values. Journal of Functional
Analysis, 244 (2007), 387--398.


\bibitem[OY12]{OY12}
\"Ostensson J., Yafaev D. R. A trace formula for differential operators of arbitrary order, A panorama of modern operator theory and related topics. – Springer Basel, 2012, 541--570.


\bibitem[PT]{PT} J. P\"oschel, E. Trubowitz, Inverse spectral
theory, Academic Press, Boston, 1987.

\bibitem  [S00]{S00} Simon, B. Resonances in one dimension and Fredholm
determinants,  J. Funct. Anal. 178 (2000), no. 2, 396--420.

\bibitem  [S05]{S05}  Simon, B. Trace ideals and their applications.
Second edition. Mathematical Surveys and Monographs, 120. AMS,
Providence, RI, 2005.


\bibitem[RS72]{RS72} Reed, M., Simon, B.
Methods of modern mathematical physics. Vol. 1.
Functional analysis, Academic Press, New York, 1972.

\bibitem[RS75]{RS75}  Reed, M.,  Simon, B.
Methods of modern mathematical
physics. II. Fourier analysis, self-adjointness, Academic Press,
New York,
1975.

\bibitem[RS79]{RS79} Reed, M.; Simon, B.  Methods of Modern Mathematical
Physics, Vol. III: Scattering Theory, Academic Press, New York, 1979.


\bibitem[RS78]{RS78} Reed, M.; Simon, B. Methods of Modern Mathematical
Physics, Vol.IV: Analysis of Operators, Academic Press, New York, 1978.


\bibitem [SZ91]{SZ91} Sj\"ostrand, J.; Zworski, M. Complex scaling and the
 distribution of scattering poles. J. Amer. Math. Soc. 4 (1991), no. 4, 729--769.

\bibitem [S90]{S90}
Sukhanov V. V. An inverse problem for a selfadjoint differential operator on the line, Mathematics of the USSR-Sbornik. – 65(1990), No 1, 249--266.

\bibitem [Y00]{Y00}
Yurko V. A. Inverse spectral problems for linear differential operators and their applications. – CRC Press, 2000.




\bibitem  [Z87]{Z87} Zworski, M. Distribution of poles for scattering
on the real line, J. Funct. Anal. 73(1987), 277--296.

\bibitem [Z02]{Z02} Zworski, M. SIAM, J.  Math. Analysis, "A remark on
isopolar potentials" 82(2002), no. 6, 1823--1826.

\bibitem [Z99]{Z99} Zworski, M.
Resonances in physics and geometry, 
Notices of the AMS 46(1999), no. 3, 319--328.






\end{thebibliography}
\end{document}